\documentclass[fleqn,usenatbib]{mnras}

\usepackage{newtxtext}
\usepackage{amsmath}
\usepackage{MnSymbol}
\usepackage{enumitem}
\usepackage{cancel}
\usepackage[T1]{fontenc}
\usepackage{longtable}
\usepackage{natbib,threeparttable,rotating,placeins}
\usepackage{tablefootnote}

\newcommand{\cppp}{C^{+3}}

\newcommand{\civ}{C~{\sc iv}}
\newcommand{\siiv}{Si~{\sc iv}}
\newcommand{\oiii}{{\sc [}O~{\sc iii}{\sc ]}}
\newcommand{\oiv}{O~{\sc iv}{\sc ]}}
\newcommand{\feii}{Fe~{\sc ii}}
\newcommand{\mgii}{Mg~{\sc ii}}
\newcommand{\femg}{Fe~{\sc ii}/Mg~{\sc ii}}
\newcommand{\fehb}{Fe~{\sc ii}/H$\beta$}
\newcommand{\hbeta}{H$\beta$}
\newcommand{\mghb}{Mg~{\sc ii}/H$\beta$}
\newcommand{\nagao}{(Si~{\sc iv}+O~{\sc iv}{\sc ]})/C~{\sc iv}}

\usepackage{CJKutf8}
\usepackage{amsmath,graphicx,longtable,hyperref}
\usepackage{ulem,changepage}
\usepackage{xcolor}
\hypersetup{
    colorlinks,
    linkcolor={red!50!black},
    citecolor={blue!50!black},
    urlcolor={blue!80!black}
}

\graphicspath{{./}{figures/}}

\title[Stellar Population in AGN Disks]{Stellar Population and Metal Production in AGN Disks}

\author[Fryer et al.]{ \thanks{E-mail: fryer@lanl.gov (CF)} 
Chris L. Fryer$^{1}$, Jiamu Huang $^{2}$, Mohamad Ali-Dib$^{3}$, Amaya Andrews$^{1}$, Zhenghao Xu$^{2}$, Douglas N. C. Lin $^{4}$
\\
$^{1}$Los Alamos National Laboratory, USA\\
$^{2}$Department of Physics, University of California, Santa Barbara, CA 93106, USA\\
$^{3}$Center for Astrophysics and Space Science (CASS), New York University Abu Dhabi, PO Box 129188, UAE\\
$^{4}$Department of Astronomy and Astrophysics, University of California, Santa Cruz, CA 95064, USA
}

\date{Accepted XXX. Received YYY; in original form ZZZ.}
\pubyear{2022}

\begin{document}
\label{firstpage}

\begin{CJK*}{UTF8}{gbsn}


\maketitle
\end{CJK*}

\begin{abstract}
As gravitational wave detections increase the number of observed compact binaries (consisting of neutron stars or blacks), we begin to probe the different conditions producing these binaries.  Most studies of compact remnant formation focus either on stellar collapse from the evolution of field binary stars in gas-free environments or the formation of stars in clusters where dynamical interactions capture the compact objects, forming binaries.  But a third scenario exists.  In this paper, we study the fate of massive stars formed, accrete gas, and evolve in the dense disks surrounding supermassive black holes.  We calculate the explosions produced and compact objects formed by the collapse of these massive stars.  Nucleosynthetic yields may provide an ideal, directly observable, diagnostic of the formation and fate of these stars in active galactic nuclei. We present a first study of the explosive yields from these stars, comparing these yields with the observed nucleosynthetic signatures in the disks around supermassive stars {with quasars}.  We show that, even though these stars tend to form black holes, their rapid rotation leads to disks that can eject a considerable amount of iron during the collapse of the star.  The nucleosynthetic yields from these stars can produce constraints on the number of systems formed in this manner, but further work is needed to exploit variations from the initial models presented in this paper.

\end{abstract}


\section{Introduction}
\label{sec:introduction}

Our understanding of compact remnants (black holes and neutron stars) and their formation mechanisms has increased in the past few years due to a rapidly growing set of gravitational wave observations probing the properties (masses and spins) of these compact objects~\citep{PhysRevX.9.031040,PhysRevX.11.021053,PhysRevX.13.041039,PhysRevD.109.022001}.  Most studies of the remnant formation have focused on field star evolution~\citep[e.g.][]{1999ApJ...526..152F,2000A&A...355..479B,2018MNRAS.481.1908K,2020A&A...636A.104B,2021MNRAS.508.5028B,2023MNRAS.525..706R}.  Other formation scenarios have been considered including star formation (and mergers) in dense stellar clusters~\citep{2016PhRvD..93h4029R,2020MNRAS.497.1563R,2021MNRAS.505.3844B} and star formation in extreme conditions such as in the disks of Active Galactic Nuclei (AGN)~\citep{1978AcA....28...91P,2016MNRAS.462.3302E,2019PhRvL.123r1101Y,2021ApJ...910..152Z,2022MNRAS.517.4034S,2022LRR....25....1M,2023Univ....9..138A,2024MNRAS.531.3479M,2024PhRvD.110d3023K}.  To fully analyze this growing set of binary compact remnant observations, we must understand all of the potential formation scenarios of neutron stars and black holes.

One of the least-studied formation scenarios is the evolution and collapse of massive stars in AGN disks.  The conditions in the standard $\alpha-$disk prescription of the accretion disks around a supermassive black hole with masses between $10^6-10^8\,M_\odot$ are such that the disk is susceptible to gravitational instabilities~\citep{1978AcA....28...91P}.  These instabilities, in turn, can drive the formation of stars within the disk.  Star formation in AGN disks has been studied for over two decades~\citep{2003MNRAS.339..937G,2004ApJ...608..108G,2005ApJ...630..167T, chen2023}.
AGN disks also capture mature stars in the nuclear clusters surrounding the central supermassive black holes (SMBHs).
\citep{artymowicz1993, davies2020, macleod2020}. 

The evolution of these stars has recently been numerically modeled using the \texttt{MESA} scheme~\citep{cantiello2021,2022ApJ...929..133J,2023MNRAS.526.5824A}. The fact that these stars form immersed in a dense disk causes them to evolve very differently than field stars.  
The relatively high-densities surrounding these stars cause them to accrete mass until the star, no matter its starting mass, exceeds a mass of a few hundred $M_\odot$.  This high mass then drives mass-loss when the star reaches its Eddington luminosity~\citep{2023MNRAS.526.5824A}.  

In this paper, we calculate the further evolution and final fate of these peculiar stars.
We assess both their potential for driving supernova-like explosions and the properties of the 
compact remnants formed in their collapse.  One way to test the formation 
rate and final fate of these stars is to study the nucleosynthetic 
yields from these stars~\citep{Huang2023}.  With our explosion calculations we can calculate the detailed yields from these AGN disk stars, combining the yields from the stellar winds and the collapse-driven explosions.  We then compare the predicted yields to those observed in AGN disks. 

The outline of the paper is as follows.  We first review the evolution of the AGN-disk stars used in this study (\S\ref{sec:stellar}).  Using the structures (e.g. density, temperature, composition profiles) from the stellar models described in \S\ref{sec:stellar}, we then determine the fate of these stars (remnant formation and explosion properties), leveraging current analyses of the collapse of field stars (\S\ref{sec:collapse}).   The yields (combining yields ejected in stellar winds, material processed in disk formed around the collapsed star and the stellar material ejected by the disk wind) from the stellar collapse are discussed in \S\ref{sec:simyields}.  These yields are compared to observations {of AGN disks in quasar systems} in \S\ref{sec:observ}.  We conclude with a summary of our results and a discussion of future work.

\section{Stellar Evolution in AGN Disks}
\label{sec:stellar}

Stellar evolution in the dense environments of AGN disks proceeds very differently than that of field stars.  The dense circumstellar AGN disk material leads to continued accretion onto the star.  Independent of their initial mass, these AGN-disk stars 
can acquire up to several hundred solar masses before they efficiently shed mass through intense stellar winds. If the accreted material in the envelope is well  mixed with the He ashes in the nuclear-burning (through CNO cycle) core, this ``metabolic'' process would enable the massive stars embedded in the dense inner regions of AGN disks to indefinitely remain on the main  sequence \citep{cantiello2021}.  However, if their radiative zone provides an effective buffer to prevent significant mixing, the He and N abundances in these stars' core would increase through the CNO cycle on the main sequence.  Consequently, their mass-luminosity function would evolve and their Eddington limit would be marginally maintained by shedding envelope mass with net enhanced He and N yields.  After nearly all the H content is exhausted in their cores and a major fraction of their total mass is lost, these stars undergo post main sequence evolution.  Subsequently, they release several $M_\odot$ of He and C+O-enriched yields before their evolution calculation with the MESA code is halted at the onset of Si burning~\citep{2023MNRAS.526.5824A}.  Shortly thereafter, the Mg and Si-laden cores collapse.  In this paper, we consider the 
possibility of type II supernovae resulting from the core collapse and determine their yields of iron peak elements.  For this first study of the compact remnants, explosions and yields from these massive stars, we simulate stellar evolution varying the rotational velocity, the metallicity, and the AGN disk density (corresponding to different positions in the AGN disk).

Our main-sequence and post-main-sequence stellar evolution models essentially follow the prescription outlined in  \cite{cantiello2021}. We adopt the modification made by \cite{2023MNRAS.526.5824A} by switching off the ``extra mixing'' in the radiative zones which was implemented by \cite{cantiello2021} to enable the replenishment of freshly-accreted H-rich gas to the core where the conversion of H to He through the CNO cycle is ongoing.  We initiate a simulation with a young stellar object with a mass $M_{\rm initial}= 1 M_\odot$ embedded in an AGN disk with nominal boundary conditions $\rho_{\rm disk} =6\times 10^{-17}$ g cm$^{-3}$ and sound speed $c_{s_{\rm disk}}= 10^6$ cm s$^{-1}$. The star initially grows through Bondi accretion. As the star's mass increases above roughly a few hundred $M_\odot$, its luminosity rapidly approaches the Eddington limit with an intensifying radiation-pressure driven wind.

{We model this wind following \cite{cantiello2021} as a super-Eddington outflow at the escape velocity with a mass loss rate set by the excess luminosity and  defined as:
\begin{equation}
    \dot{M}_{\mathrm{Edd}}=-\frac{L_*}{v_{\mathrm{esc}}^2}\left[1+\tanh \left(\frac{L_*-L_{\mathrm{Edd}}}{0.1 L_{\mathrm{Edd}}}\right)\right]
\end{equation}
where the tanh term is to help numerical convergence.}

A balance between 
the accretion and wind mass-loss rates is established with a mass $M_{\rm eq} 
\sim 630 M_\odot$.  We take into account the partial retention of the  
He-laden wind and the local recycling of the accreted gas due to the 
inflow-outflow congestion\citep{chen2024b}.  Provided there is no ``extra'' 
rotational mixing, the presence of a radiative envelope in main sequence  stars
maintains a barrier 
which cuts off the H replenishment to the nuclear furnace.  As He increases 
with H depletion in its isolated core, the star brightens 
and the state of marginal Eddington-limited equilibrium is maintained through 
the radiation-driven mass reduction.  In comparison with the disk gas, the 
gas released by the main sequence stars carries excess He and N yields with 
slightly C+O depletion (because the C+N+O abundance is conserved through 
the CNO burning). After the H abundance vanishes in its core, the star's 
mass is reduced to $\sim 25-30 M_\odot$ and it transitions to post main 
sequence evolution with the onset of He burning.  For a more detailed 
description of the approaches, we refer the reader to the work of 
\cite{cantiello2021} on the immortal and \cite{2023MNRAS.526.5824A} 
on the metamorphic stars.

Our previous models neglect the effect of stellar rotation. In principle,
the freshly-accreted H-rich gas may pass through the radiative envelope 
via meridional circulation and mix with the He-ashes-laden gas in the core.  But, 
unless the star is spinning at a large fraction of the break-up rate, 
limited elemental diffusion cannot sufficiently replenish the hydrogen 
fuel to significantly 
prolong the CNO cycle on the main sequence.  Carrying a maximum amount 
of Keplerian specific angular momentum at the stellar surface, the 
accreted gas is supplied by the gravito-turbulent eddies in AGN 
disks.  On scales up to the disk thickness (larger than the star's 
Bondi radius), the eddies' spin vector sporadically re-orients on their 
turn-over timescale (comparable to the disk's orbital timescale 
$\tau_{\rm eddy} = \Omega^{-1}$). Over the much longer characteristic 
accretion time scale ($M_\star/ {\dot M}_\star$), the stochastic accretion 
from the randomly aligned eddies reduces the net angular-momentum deposition 
onto the star well below its rotational break-up range \citep{chenlin2023}.    
Moreover, angular momentum may be removed from the star by its outflowing 
wind due to the coupling between different regions of the stellar interior by convection or magnetic fields.  

Based on these considerations, we neglect the effect of stellar spin in
inducing ``extra'' rotational mixing during AGN stars' main sequence 
evolution.  However, during the star's brief post main sequence evolution, 
there are three effects which can significantly boost angular momentum 
acquisition -- especially during the super giant phase when its mass has 
reduced to $M_\star \lesssim 25-30 M_\odot$,
shock radius $R_{\rm rot}$ to $\sim 2 R_\odot$,
and its photosphere radius has expanded 
to $R_{\rm ph, rot} \sim 10^2 R_\odot$.  First, with a rotational velocity $v_{\rm rot}$
comparable to the Keplerian speed at $R_{\rm ph, rot}$,
$v_{\rm rot} \sim V_{\rm Kep} (R_{\rm ph, rot}) \sim 200$ km s$^{-1} \sim 0.1 V_{\rm Kep} (R_{\rm rot})$ 
and the specific angular momentum $j_\star \simeq (G M_\star R_{\rm rot})^{1/2} 
\sim 2 \times 10^{19}$ cm s$^{-1}$ on the star's shock radius.  Second, 
the stellar envelope is fully convective.  Although H, He, and heavy elements 
are well mixed even without any rotational effects, the accretion rate of disk 
gas is inadequate to replenish either H or He to slow-down the star's post-main-sequence 
evolution. Finally, the star's evolution time scale is reduced to $\sim 10^{3-4}$ yr 
which is comparable to the turnover timescale $\tau_{\rm eddy}$ of gravito-turbulent 
eddies (which is comparable to the star's orbital period around the SMBH).  Thereafter, 
the accreted gas has no time to undergo spin reorientation and 
the rotation speed of the star $v_{\rm rot}$ may have magnitudes  range from negligible 
to small, $\sim {\mathcal O} (0.1$), fraction of $v_{\rm Kep} (R_{\rm rot})$.  In 
principle, super Keplerian rotation speed may also be temporarily established near 
the star's surface after some extreme merger events \citep{chenlin2024} although it 
would rapidly spin down through intense mass and angular momentum losses.  

A non-negligible amount of spin angular momentum for the post-main-sequence stars
introduces the possibility of disk formation around compact remnants of collapse 
core (\S\ref{sec:collapse}).
In order to take these effects into account, we consider a range 
of rotation speed $v_{\rm rot}$ on the stellar surface. 
We utilize the \texttt{new\_surface\_rotation\_v} 
and \texttt{adjust\_J\_q\_limit} flags in the \texttt{MESA} code to compute the star's 
internal specific angular momentum distribution for these 
outer boundary conditions. For the fiducial models (v5, v20, v100, and v200),
we adopt solar composition for the disk gas.  We also include additional variations 
(v200lowden and v200lowZ) for low disk density ($=10^{-17}$g cm$^{-3}$) and 
purely H and He gas supply 
\citep{cantiello2021} in the v200lowden and v200lowZ models respectively.  
A comprehensive evaluation of the star's most likely rotational 
properties will be examined elsewhere.

With mass $M_\star=M_{\rm rot} = 25-30 M_\odot$, shock radius $R_{\rm rot} \sim R_\odot$ (Eq. 33 in \citet{cantiello2021}),
radius of photosphere $R_{\rm ph, rot} 
\simeq 120 R_\odot$, rotation speed at the stellar surface $v_{\rm rot} = 5-200$ 
km s$^{-1}$ (ranging negligible to modest
fraction of $v_{\rm Kep}$), and
total angular momentum $J_{\rm rot}$ (Table~\ref{tab:star}),  we relax 
the rotational models
subjected to the influence of magnetic coupling to obtain the angular 
momentum distribution within $R_{\rm rot}$ \citep{paxton2013} at the 
end of main-sequene of evolution. We continue the \texttt{MESA} calculation 
of the star's post-main-sequence evolution until the onset of Si burning.  
We use the standard \texttt{MESA} algorithm to compute the evolution 
of star's angular momentum and composition distribution. 
Radiation-pressure driven mass loss (mostly in He, C, and O, 
\cite{2023MNRAS.526.5824A}) 
along with angular momentum removal, enables the star's luminosity 
to be maintained at a marginal Eddington-limited level.  

\begin{figure}
\centering
\includegraphics[width=0.47\textwidth]{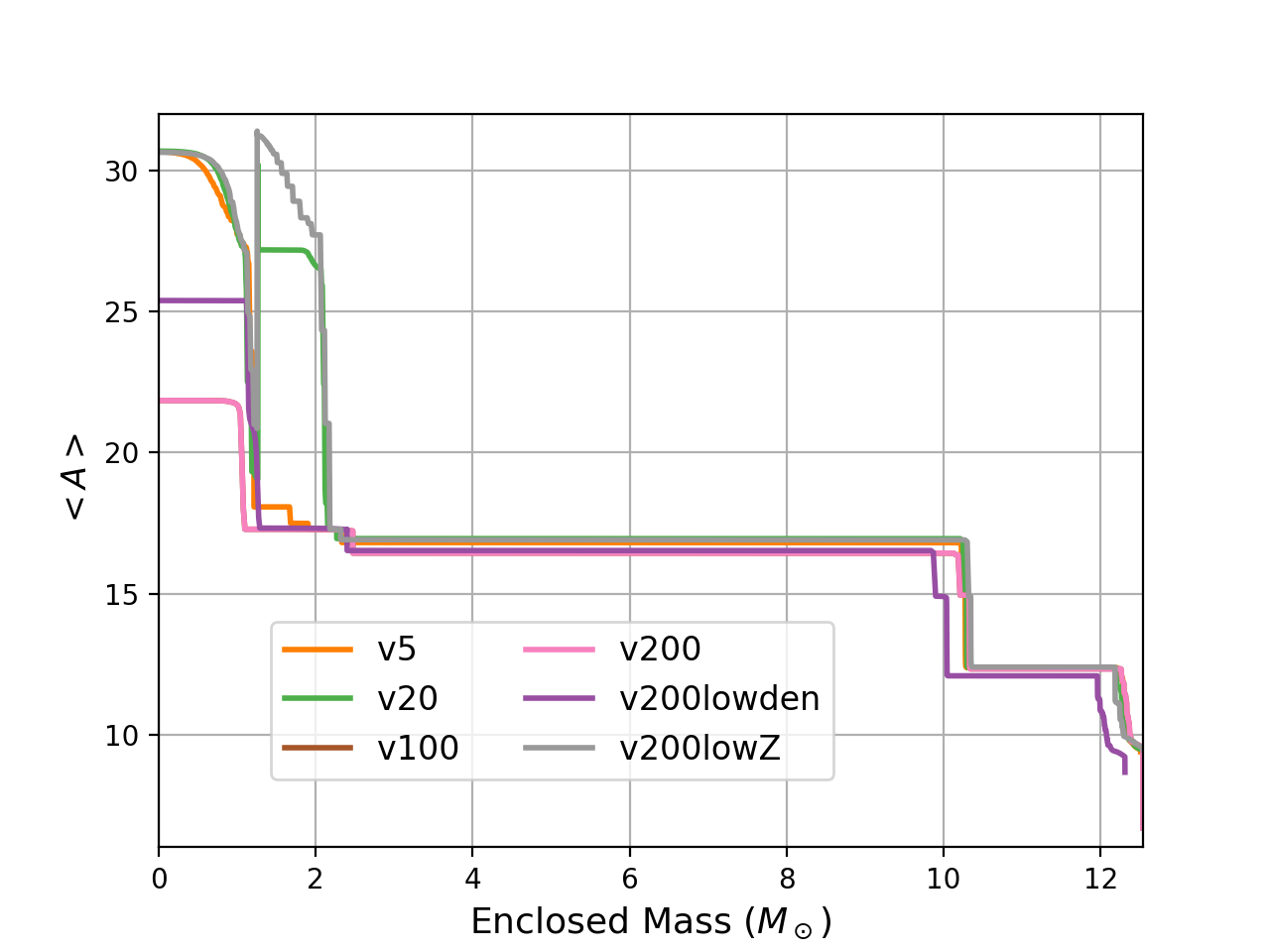}
\caption{Average atomic mass of our stars at collapse as a function of the enclosed mass coordinate.  We vary the rotation speed to include outer star spin-up from 5-200\,${\rm km \, s^{-1}}$.  
We also include a model with embedded in a lower metallicity disk and a model at lower metallicity - see Table~\ref{tab:BHmass}.  The variation in the inner 2\,M$_\odot$ is due to slight differences in the final time dump of the simulations as the material is burning to Silicon.  The sharp jumps in abundances at $1\,M_\odot$ occur at the dividing line between core and shell C/O burning.  As we evolve the star further, these abundances would become more smooth.  Beyond those differences, the models are very similar.  This is because, in all models, the stars accrete material until they achieve the same high mass before losing most of the mass through winds.  
{In our case, the wind mass loss is not metallacity-driven as the Eddington luminosity is set by the electron scattering opacity.} These stars end up with very large O cores that will collapse to black holes in all supernova models. }
\label{fig:abar}
\end{figure}

\begin{figure}
\centering
\includegraphics[width=0.47\textwidth]{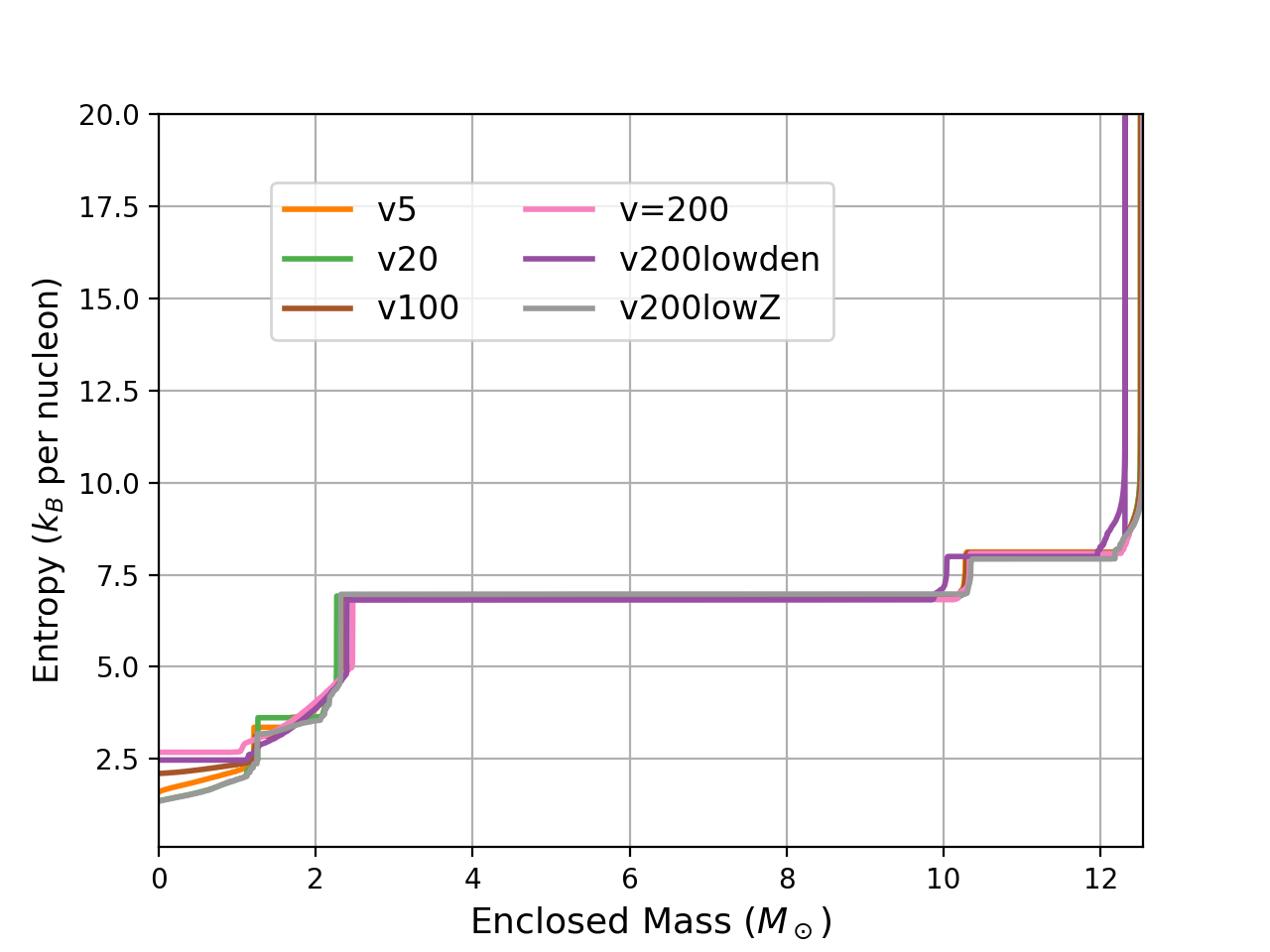}
\caption{Entropy versus enclosed mass for the same suite of models as Figure~\ref{fig:abar}/Table~\ref{tab:BHmass}.  The entropy of all our stars are nearly identical.}
\label{fig:ent}
\end{figure}

The \texttt{MESA} calculation is terminated at the onset of 
Si burning with a temperature $10^{9.5}$ K before the 
core collapse.  The star has a mass $M_{\rm fi}=12 M_\odot$, a radius $R_{\rm fi}$, and
a total angular momentum $J_{\rm fi}$ 
(Table~\ref{tab:star}).  In all models, C and O contribute to 
most of the mass and a modest mass $M_{\rm Si} \sim 1-2 M_\odot$ 
at the center of the pre-collapsing core have $> 10\%$ concentration
of Silicon (Figure~\ref{fig:abar}). Only a negligible fraction of
Si and Mg yield is released to the disk before the collapse. 
The corresponding entropy 
profiles are shown in Figure~\ref{fig:ent}.  Just as with the 
abundances, there is very little variation in entropy in these models.  The amount of mass
reduction ($M_{\rm fi}/M_{\rm rot} \sim 0.5$)
does not depend on $v_{\rm rot}$ which indicates
that the mass loss is mostly driven by radiation pressure, not by centrifugal force.  

Within the shock radius $R_{\rm fi}$  
\citep{cantiello2021}, much smaller than
the photospheric radius $R_{\rm ph, fi}$, 
there are noted differences in the specific angular momentum
distribution (Figure~\ref{fig:rot}) between the three rotational models 
(with surface $v_{\rm rot}=5, 20,$ and 200 km s$^{-1}$ in Table \ref{tab:star}).
One major uncertainty in the \texttt{MESA} models \citep{paxton2013} is the efficiency of angular 
momentum transfer between different layers of the envelope and 
convective/radiative interface (due to transition in the composition 
in Figure~\ref{fig:abar}). Magnetic breaking by the stellar wind on star's
surface and the evolution of internal density distribution (changes in 
the envelope size) can lead to differential rotation. In principle, 
strong coupling can re-establish uniform rotation with specific angular 
momentum proportional to the square of distance from the stellar core.  
Here, we use the standard \texttt{MESA} prescription for partial 
angular-momentum diffusion between layers with different composition.  
At interfaces where convection is stabilized by molecular weight gradient
there are significant drops in the angular velocity (Figure \ref{fig:rot}).  The 
difference  between 
$M_{\rm rot}/M_{\rm fi} \sim 2$
and $J_{\rm rot}/ J_{\rm fi} \sim 3$
for models v5, v20, and v200 (also
between $M_{\rm rot}/M_{\rm fi} \sim 2$
and $J_{\rm rot}/ J_{\rm fi} \sim 8$ for 
the v200lowZ and v200lowden models in Table 
\ref{tab:star}) 
is a manifestation of a modest amount of 
magnetic breaking imposed by the wind.

\begin{table*}
\begin{center}
\begin{tabular}{l|cccccccccc}
\hline\hline              
Model & $M_{\rm rot}$ & $v_{\rm rot}$ & $R_{\rm rot}$ & $J_{\rm rot}$ & $M_{\rm fi} \simeq M_{\rm CO}$ & $M_{\rm Si}$ & $R_{\rm fi}$ & $R_{\rm ph, fi}$ & $J_{\rm fi}$ \\
 & ($M_\odot$) & ($10^7 {\rm cm \, s^{-1}}$) &  $(R_\odot)$ &($10^{52} {\rm g \, cm^2 \, s^{-1}}$) & ($M_\odot$) & ($M_\odot$) & ($R_\odot$) & $(R_\odot)$ & ($10^{52} {\rm g \, cm^2 \, s^{-1}}$) \\
& & & \\
\hline
v5 & 25 & 0.05 & 1.28 & 0.014 & 12.5 & 1.9 & 
0.38 &303.4 & 0.0045  \\
v20 & 25 & 0.2 & 1.28 & 0.060 & 12.5 & 2.1 & 0.39 & 303.4 & 0.018 \\
v100 & 25 & 1 & 1.28 & 0.29 & 12.3 & 1.3 & 0.37 & 304.04  & 0.089 \\
v200 & 25 & 2 & 1.29 & 0.59 & 12.5 & 1.1 & 0.36 & 305.6  & 0.18 \\
v200lowZ & 29 & 2 & 1.39 & 1.71 & 12.5 & 2.2 & 0.38 & 306.1 & 0.219 \\
v200lowden & 30 & 2 & 1.34 & 4.38 & 12.3 & 1.3 & 0.39 & 8.41 & 0.498 \\
\hline\hline
\end{tabular}
\caption{Stellar Models with AGN-disk Accretion.  The total pre-collapse stellar mass
$M_{\rm fi}$ is mostly that $M_{\rm CO}$ of C and O.  The silicon core is the region 
where the silicon mass 
fraction is greater than 10\%. It contains a mass $M_{\rm Si}$. }
\label{tab:star}
\end{center}
\end{table*}

\section{Stellar Collapse:  Remnant Masses, Ejecta Properties and Gravitational Waves}
\label{sec:collapse}

Although we do not, and cannot, follow the evolution of the stars to collapse, 
we can infer their fate based on the structure near carbon exhaustion
and the onset of Si burning with the \texttt{MESA} models.  
A number of prescriptions exist to 
determine the compact remnant produced in stellar collapse~\citep{2012ApJ...749...91F,2022ApJ...931...94F}.  For all of these 
prescriptions, we do not expect a supernova explosion (even a weak 
supernova explosion) for slowly or non rotating stars with CO core 
masses $(M_{\rm CO}) \geq 8\,M_\odot$ at the onset of
Si burning.  As shown in Table~\ref{tab:star}
and Figure \ref{fig:abar}, all of the stars produced in AGN disks 
produce CO cores above this mass.  We expect the residual core of these 
stars to collapse and form black holes.

In the absence of a sufficiently large spin angular momentum, these stars are unlikely to eject much mass during the collapse.  Under this condition, the remnant's mass would equal the mass of the collapsing star and the yields would be limited to the stellar-wind mass loss from the star (in elements lighter than Si) during their main sequence and post main sequence evolution \citep{2023MNRAS.526.5824A}.  In this section, we focus on the mass ejection from disks formed in the collapse of rotating stars.  

\subsection{Formation of disks around remnant black holes}
As we have discussed above, AGN stars likely to spin up at the end of their post main-sequence evolution as they exchange mass with the disk around the SMBH.  For our progenitors, the fast-spinning model (with $v_{\rm rot} = 200$ km s$^{-1}$) has sufficient angular momentum to produce a disk outside the remnant's gravitational radius during the
collapse.  We focus on this type of progenitors for our study.  

\begin{figure}
\centering
\includegraphics[width=0.47\textwidth]{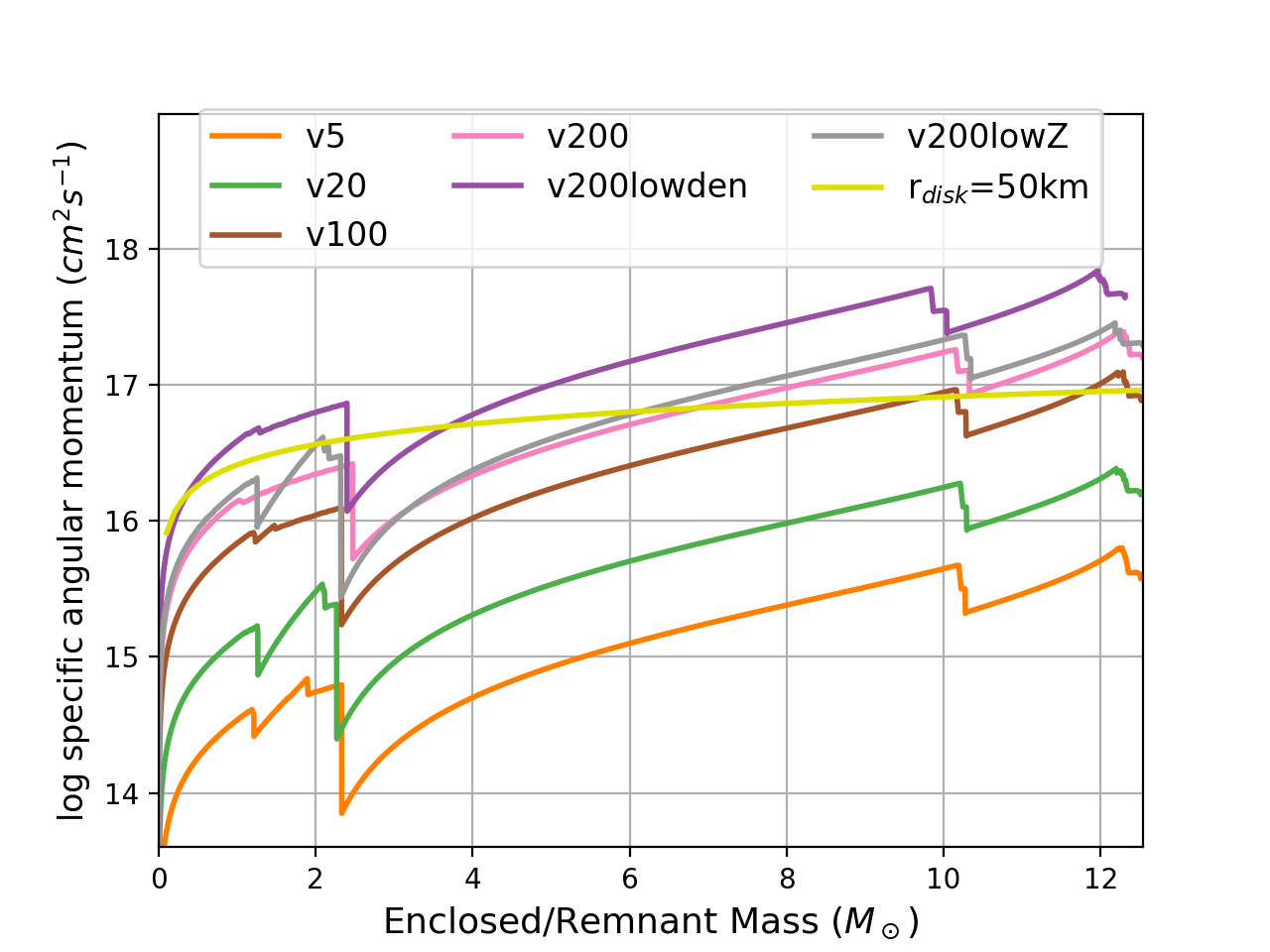}
\caption{Specific angular momentum profiles for the models described in Table~\ref{tab:BHmass}.  High angular momenta are needed to form a disk with a radius 
of 50 km.  For comparison, the angular momentum needed to form a $r_{\rm disk} \simeq$
50\,km disk is plotted as a function of the compact remnant mass $M_{\rm rem}$
set to the enclosed 
mass ($r_{\rm disk} > r_{\rm ISCO}$, Eqs. \ref{eq:rdisk} \& \ref{eq:rISCO}).  
Only our models with initial rotation velocities above 200\,${\rm km \, s^{-1}}$ have sufficiently high
angular momenta to form a disk.}
\label{fig:rot}
\end{figure}

The initial collapsing material (from the central region of the core) 
forms the black hole.  At first, the angular momentum is too low 
to significantly impede the infall and the imploding stellar material 
passes through the event horizon, adding momentum and mass to the 
remnant.  Even if the compact remnant is a neutron star, at these 
accretion rates, neutrino cooling rapidly removes the entropy from 
the material piling up on the neutron star and it quickly compresses 
becoming part of the remnant~\citep{1996ApJ...460..801F}.   As the outer, higher angular-momentum material falls onto the compact remnant, centrifugal forces cause it to stall and accumulate in the disk.  It is believed that viscous heating in the disk drives winds that eject both mass and angular momentum~\citep[see review in][]{2023ApJ...956...71K}.  As the material in the disk loses angular momentum, it moves inward.  In our model, the angular momentum deposited onto the black hole by this disk material is limited to its angular momentum at the event horizon.  

\begin{figure}
\centering
\includegraphics[width=0.47\textwidth]{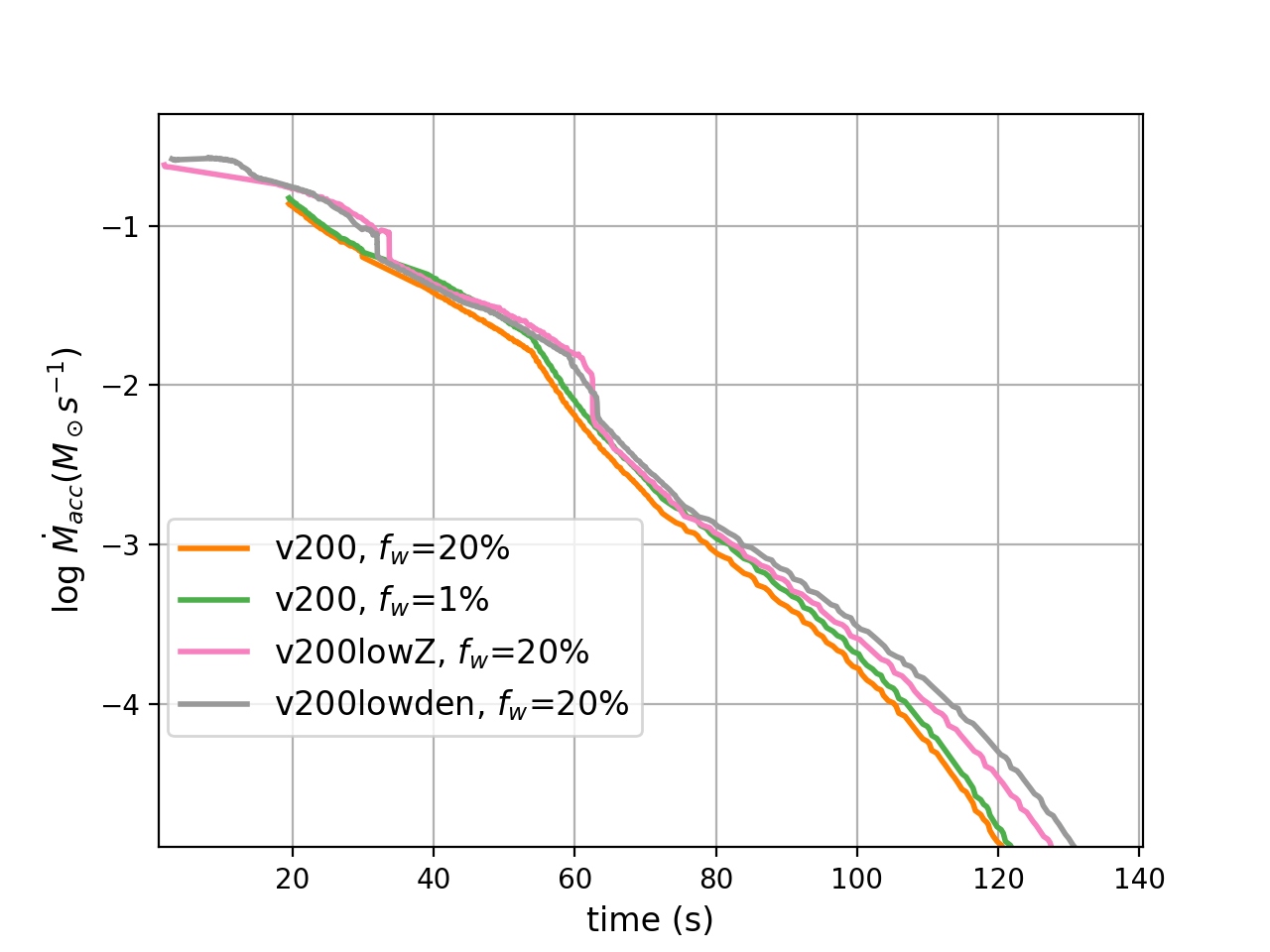}
\caption{Accretion rate onto the black hole as a function of time for the models described in Table~\ref{tab:BHmass} using the free-fall time of the collapsing stellar material (Eq. \ref{eq:tfreefall}) and an $\alpha=0.01$ accretion disk\citep{deng2020}  These accretion rates are similar  to those seen in higher-mass collapsar models~\citep{1999ApJ...518..356P}.}
\label{fig:mdot}
\end{figure}

The imploding star either accretes onto the compact object or, for sufficiently high angular momentum material, feeds the disk at roughly the free-fall time~\citep{1996ApJ...460..801F}.  The accretion rate/disk-feeding rate ($\dot{M}_{\rm disk}$) can be determined using the stellar structure:
\begin{equation}
    \dot{M}_{\rm disk} = \Delta M/\Delta t_{\rm ff} = (M_{\rm k+1} - M_{\rm k})
    /(t_{\rm ff_{\rm k+1}}-t_{\rm ff_{\rm k}})
\end{equation}
where $M_{\rm k}$ is the enclosed mass at the radial coordinate $r_{\rm k}$ 
and $t_{\rm ff_{k}}$ is the free-fall time at that coordinate:
\begin{equation}
    t_{\rm ff_{\rm k}} = \pi r_{\rm k}^{1.5}/(2 G M_{\rm BH_k})^{1/2}
    \label{eq:tfreefall}
\end{equation}
where $G$ is the gravitational constant and $M_{\rm BH_k}$ is the black hole mass.  $M_{\rm BH_k} \approx M_{\rm k}$ but will be slightly smaller due to wind mass loss (we include this mass loss in our calculations).  For low-angular momentum material, this is the accretion rate onto the black hole.  Conservation of angular momentum allows us to set the position of the disk ($r_{\rm disk}$):
\begin{equation}
    r_{\rm disk} = j_{\rm k}^2/(G M_{\rm BH_k})
\label{eq:rdisk}
\end{equation}
where $j_{\rm k}$ is the specific angular momentum in the star at position $k$.  If $r_{\rm disk}$ is greater than the radius of the innermost stable orbit ($r_{\rm ISCO}$), a disk is formed.  We assume the prograde calculate of $r_{\rm ISCO}$:
\begin{equation}
    r_{\rm ISCO} = G M_{\rm k+1}/c^2 (3 + Z_2 - \sqrt{(3-Z_1)(3+Z_1+2Z_2)})
    \label{eq:rISCO}
\end{equation}
where $Z_1=1+\sqrt[3]{1-\chi^2}(\sqrt[3]{1+\chi^2}+\sqrt[3]{1-\chi^2})$, $Z_2=\sqrt{3 \chi^2 + Z_1^2}$
and $\chi=c J_{\rm BH_k}/(G M_{\rm BH_k}^2)$ are dimensionless parameters, 
$J_{\rm BH_k}$ is the total angular momentum of the black 
hole when material at coordinate $k$ is in the disk.  
Our models with initial rotation velocities above 100\,${\rm km \, s^{-1}}$ have sufficiently high angular momenta to form a disk (Figure \ref{fig:rot}).
Since the accretion time through a compact
disk (assuming an $\alpha$-disk)~\citep{shakura1973} is typically 
shorter than the infall time from 
$R_{\rm fi}$, we can estimate the accretion rate onto the black hole as a function of time for our stellar models (Figure~\ref{fig:mdot}).

\subsection{Outflow from the disk}
\begin{figure}
\centering
\includegraphics[width=0.47\textwidth]{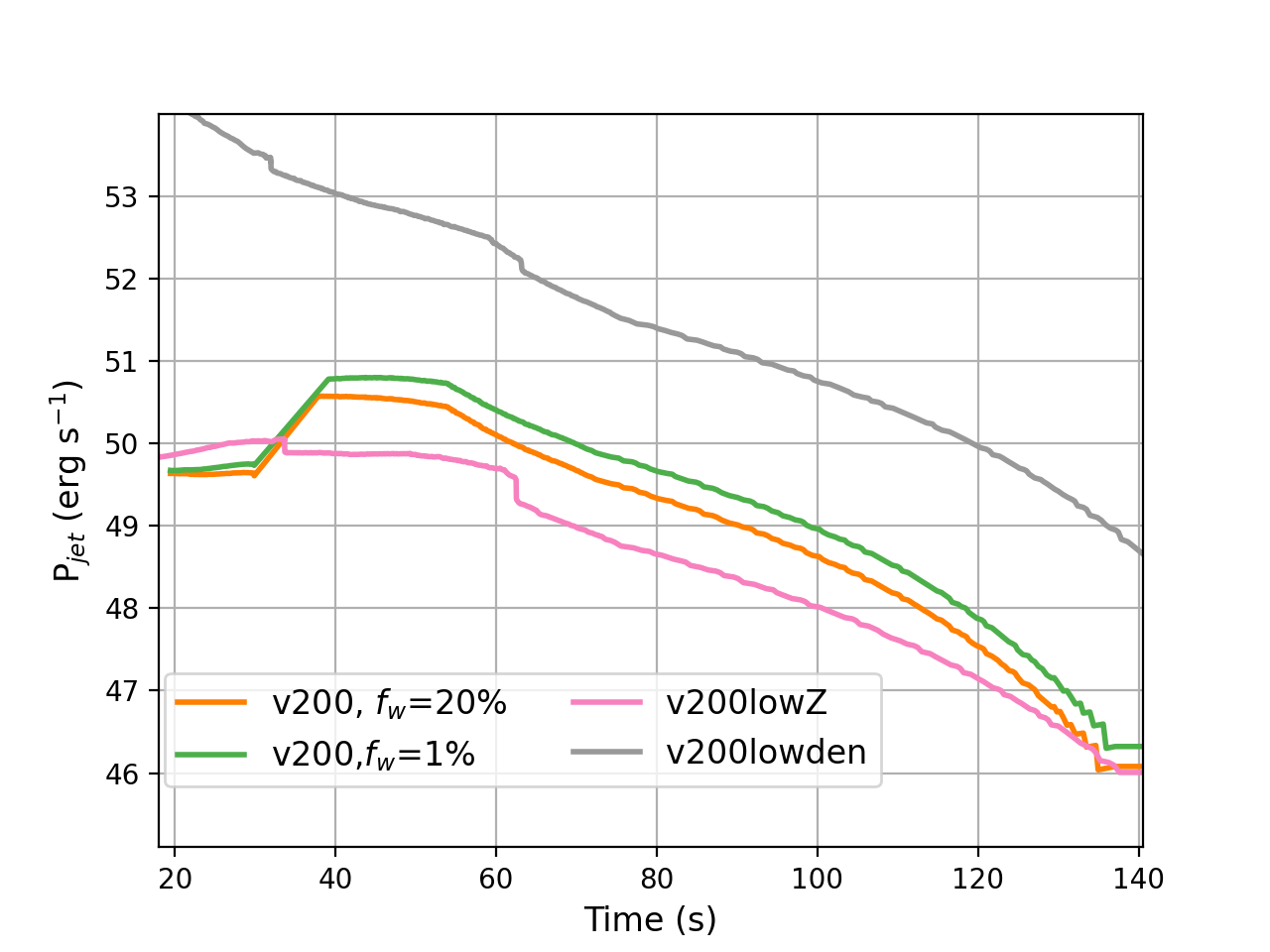}
\caption{Power of the jet (Eq. \ref{eq:lgrb})
as a function of time for the models in Table~\ref{tab:BHmass}.  
The accretion disks produced powerful jets that can reproduce the most powerful GRBs observed.  But these jets will have to propagate through the AGN disk and will be far more baryon-loaded than normal GRBs.  It is possible that these jets will eject the stellar material as the pressure wave in the jet wraps around the star.}
\label{fig:jet}
\end{figure}

The disk formed in this implosion can power outflows, both through a magnetic-dynamo driven jet~\citep{1999ApJ...518..356P} and viscosity-driven winds (also probably due to magnetic field processes).  Although neither of these outflows are fully understood, current simulations, analyses and observations all provide some insight into the behavior of these outflows.  For the jet power for our models, we use the results from~\cite{2003ApJ...591..288H}:
\begin{equation}
P_{\rm jet} =  10^{50} a_{\rm spin}^2 10^{0.1/(1-a_{\rm spin})-0.1} \left( \frac{M_{\rm BH}}{3 M_\odot} \right)^{-3} \frac{\dot{M}_{\rm disk}}{0.1M_\odot/s} {\rm \, erg \, s^{-1}}
\label{eq:lgrb}
\end{equation}
where $M_{\rm BH}, a_{\rm spin}$ are the black hole mass and spin respectively, 
${\dot M}_{\rm disk}$ is the accretion rate. 
Other models argue for a much stronger jet~\citep{2023ApJ...952L..32G}.  Figure~\ref{fig:jet} Shows the jet power using equation~\ref{eq:lgrb} as a function of time since the formation of the disk and hence jet.  The high angular momentum in the low density model produces a rapidly spinning black hole and early disk formation.  Under our jet power prescription, this produces a very powerful jet.

The jet sweeps up very little mass ($< 10^{-4}\,M_\odot$) and, in of itself, it does not alter the final remnant mass.  But as the jet punches through the star, it drives a pressure wave that can eject the outer layers of the star~\citep{1999ApJ...524..262M}.  Some fraction of the jet energy is converted into this pressure wave, driving an explosion (we assume 10\% of the jet power goes into driving this pressure wave).

\begin{table*}
\begin{center}
\begin{tabular}{l|ccccccc}
\hline\hline              
Model & $M_{\rm remnant}$ & ${M}_{\rm disk}$ & $E_{\rm jet}$ & $E_{\rm disk-wind}$ \\
 & ($M_\odot$) & ($M_\odot$) & ($10^{52}\,{\rm erg}$) &   ($10^{49}\,{\rm erg}$) \\
\hline
v5 & 12.5 & 0 & 0 & 0 \\
v20 & 12.5 & 0 & 0 & 0 \\
v100 & 12.5 & 0 & 0 & 0 \\
v200 & 11.7 & 7.7 & 1-2 & 50-100 \\
v200lowZ & 11.9 & 6.0 & 0.25-0.5 & 5-10 \\
v200lowden & 11.3 & 9.3 & 500-1000 & 5-10 \\

\hline\hline
\end{tabular}
\caption{\texttt{MESA} 
Final mass of the remnant ($M_{\rm remnant}$),
amount of mass processed through the disk
($M_{\rm disk}$), total amount of energy
released by the jet ($E_{\rm jet}$), and
from the disk wind ($E_{\rm disk-wind}$)
inferred from our collapse models.}
\label{tab:BHmass}
\end{center}
\end{table*}

\begin{figure}
\centering
\includegraphics[width=0.47\textwidth]{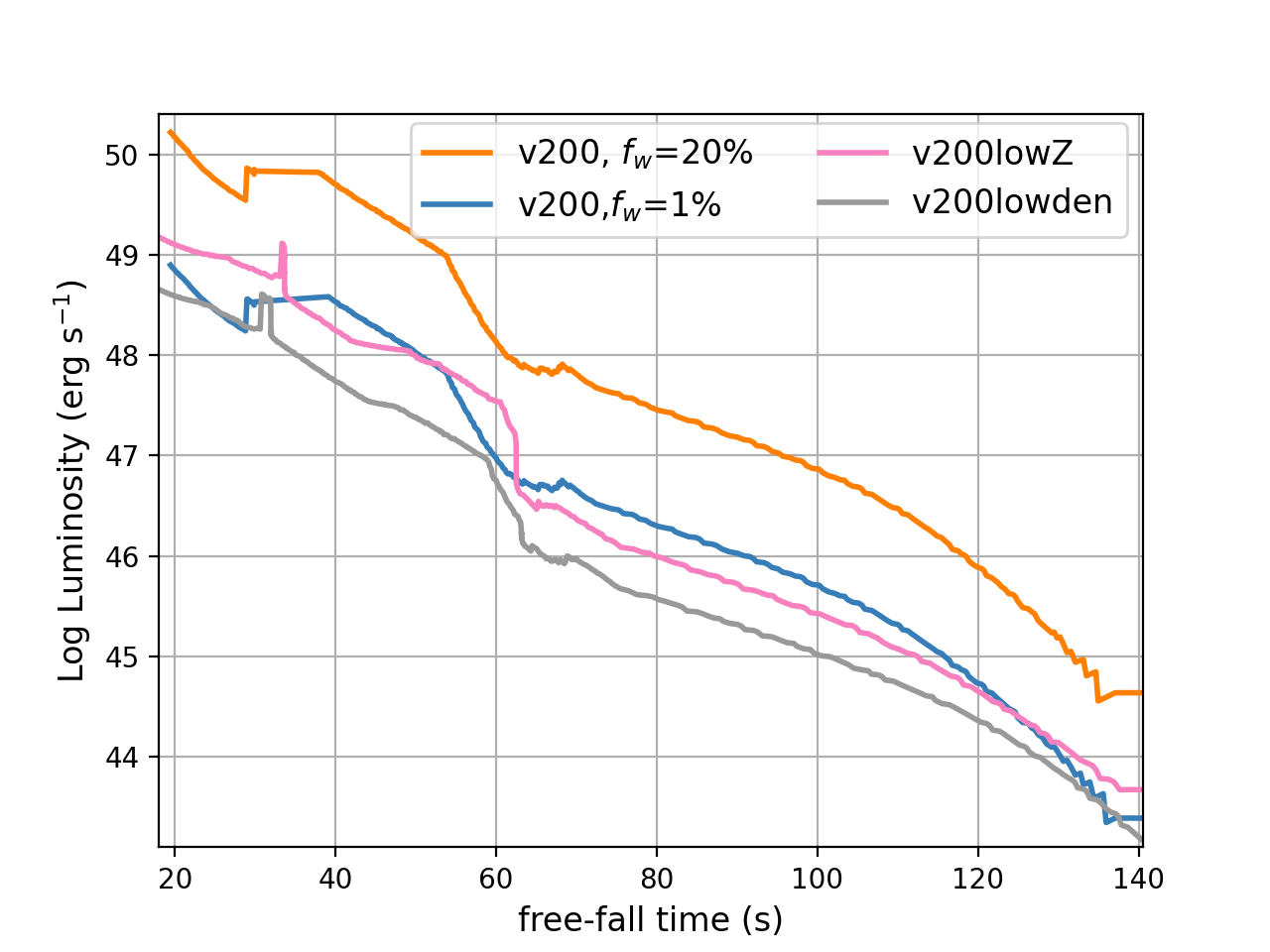}
\caption{Power of the wind outflows assuming 20\% of the disk mass is ejected with an excess energy set to 1/2 the escape velocity.  These winds will further contribute to ejecting the star.}
\label{fig:jetwind}
\end{figure}

The accretion disk itself drives an outflow whose mass is predicted to be anywhere between 1-30\% of the disk mass~\citep{2023ApJ...956...71K}.  For our yields models, we assume that 20\% of the disk mass is ejected.  We assume that the wind ejecta is uniform across the extent of the disk and the ejecta velocity is set to 1/2 of the escape velocity.  Because these outflows have more mass, they produce a more powerful explosion (Figure~\ref{fig:jetwind}) and lower-mass black holes (Table~\ref{tab:BHmass}).  

\subsection{From neutron-star to black hole cores}
In stellar collapse, the core initially forms a neutron star.  In our models, the accretion rate is high and this proto-neutron star ultimately collapses to a black hole (within a few seconds).  This 
short-lived neutron-star phase can be probed by the thermal neutrino and gravitational wave emission.  As with normal stellar-collapse, the neutrino and gravitational wave signal is limited to events in the Milky Way.  For the collapse of a rotating core, the fate depends on a parameter $\beta \equiv T/|W|$ where $T$ is the rotational energy and $|W|$ is the absolute value of the gravitational binding energy.  For $\beta>0.14$, secular instabilities can develop.  For $\beta>0.27$, dynamical instabilities develop~\citep{2002ApJ...565..430F}.  For our slower rotating stars (v5,v20) no instabilities develop.  Even for our fastest spinning cores, the instabilities do not develop until after the collapsed-core mass exceeds $\sim$1\,M$_\odot$.  This means that the core forms before instabilities occur and the collapsing core will not fragment.  However, for these fastest-spinning models (v200 series), the rotation is sufficient to develop  bar modes and gravitational waves from such stars could be detectable out to the Virgo cluster~\citep{2002ApJ...565..430F}.

\section{Yields:  Comparison of AGN stars to Field Stars}
\label{sec:simyields}

\begin{figure}
\centering
\includegraphics[width=0.47\textwidth]{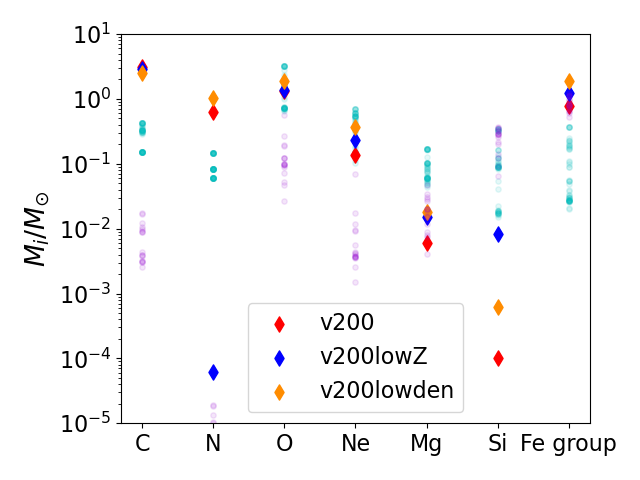}
\caption{Ejected masses of the elements for all our models.  The diamonds represent our different AGN-disk stars.  The purple dots correspond to thermonuclear supernovae and the cyan dots represent core-collapse supernovae.}
\label{fig:ejectedmasses}
\end{figure}

We have shown that the peculiar evolution of AGN-disk stars lead to stellar structurse (Section~\ref{sec:stellar}) and post-collapse explosion properties (Section~\ref{sec:collapse}) that are very different from field stars.  The nucleonsynthtetic yields from these stars probe these differences and identify distinguishing features of our AGN stars.  In this section, we discuss the yields from our AGN stellar models, contrasting them to supernova yields.  We can then compare these yields with the observed abundances in AGN disks.  As we shall discuss in Section~\ref{sec:observ}, the observed yields are limited to only a few elements:  C, N, O, Ne, Mg, Si, and Fe.  For this study, we will focus on these elements.  A summary of our calculated yields is shown in Table~\ref{tab:yields} and Figure~\ref{fig:ejectedmasses}.

\begin{table*}
\begin{center}
\begin{tabular}{l|ccccccc}
\hline\hline              
Model & C & N & O & Ne & Mg & Si & Fe peak \\
 & ($M_\odot$) & ($M_\odot$) & ($M_\odot$) & ($M_\odot$) & ($M_\odot$) & ($M_\odot$) & ($M_\odot$) \\
\hline
Stellar Yields & & & & & & & \\

v200 & 2.95 & 0.63 & 0.86  & $1.55 \times 10^{-2}$ & $2.70 \times 10^{-5}$ & $4.0 \times 10^{-6}$ & $8.0 \times 10^{-6}$\\
v200lowZ & 2.73 & $6.2 \times 10^{-5}$ & 0.67 & $1.1 \times 10^{-4}$ & $3.1 \times 10^{-9}$ & 0 & 0 \\
v200lowden & 2.22 & 1.04 & 0.75 & $1.02 \times 10^{-2}$ & $4.4 \times 10^{-5}$ & $7.0 \times 10^{-6}$ & $1.4 \times 10^{-5}$\\

\hline
Disk Yields & & & & & & & \\
v200 & $6.14 \times 10^{-9}$ & $8.50 \times 10^{-13}$ & $1.67 \times 10^{-8}$ & $8.73 \times 10^{-10}$ & $5.01 \times 10^{-10}$ & $1.96 \times 10^{-8}$ & 0.77\\
v200lowZ & $3.52 \times 10^{-8}$ & $7.25 \times 10^ {-12}$ & $9.19 \times 10^{-8}$ & $6.70 \times 10^{-9}$ & $ 3.98 \times 10^{-9}$& $1.48 \times 10^{-7}$ & 1.23 \\
v200lowden & $2.88 \times 10^{-5}$ & $2.55 \times 10^{-11}$ & $1.78 \times 10^{-4}$ & $1.56 \times 10^{-5}$ & $1.57 \times 10^{-5}$ & $2.50 \times 10^{-4}$ & 1.87 \\
\hline
Disruption Yields & & & & & & & \\
v200 & 0.142 & 0.0 & 0.476 & 0.120 & 0.00589 & $9.92 \times 10^{-5}$ & 0.0 \\
v200lowZ & 0.186 & 0.0 & 0.723 & 0.231 & 0.0149 & 0.0084 & 0.0 \\
v200lowden & 0.273 & 0.0 & 1.172 & 0.363 & 0.0180 & 0.00036 & 0.0 \\
\hline\hline
Field Stars & & & & & & & \\
\hline

Thermonuclear SNe  & & & & & & & \\
 N1 & 0.00261 & $2.92 \times 10^{-6}$ & 0.0263 & 0.00153 & 0.00402 & 0.0641 & 1.25 \\
 N3 & 0.00990 & $9.94 \times 10^{-6}$ & 0.0474 & 0.00370 & 0.00736 & 0.0914 & 1.17 \\
 N5 & 0.00905 & $8.47 \times 10^{-6}$ & 0.563 & 0.00404 & 0.00875 & 0.121 & 1.12 \\
 N10 & 0.00443 & $3.85 \times 10^{-6}$ & 0.0516 & 0.00253 & 0.00789 & 0.140 & 1.10 \\ 
 N20 & 0.00920 & $8.35 \times 10^{-6}$ & 0.0904 & 0.00567 & 0.0149 & 0.201 & 0.952 \\ 
 N40 & 0.00390 & $4.31 \times 10^{-6}$ & 0.0989 & 0.00421 & 0.0156 & 0.262 & 0.855 \\ 
 N100H & 0.00387 & $4.26 \times 10^{-6}$ & 0.0730 & 0.00374 & 0.0117 & 0.214 & 0.961 \\ 
 N100 & 0.00304 & $3.21 \times 10^{-6}$ & 0.101 & 0.00357 & 0.0154 & 0.287 & 0.815 \\ 
 N100L & 0.00385 & $3.98 \times 10^{-6}$ & 0.124 & 0.00439 & 0.0185 & 0.359 & 0.644 \\ 
 N150 & 0.0172 & $1.84 \times 10^{-5}$ & 0.124 & 0.00920 & 0.0198 & 0.275 & 0.781 \\ 
 N200 & 0.0121 & $1.33 \times 10^{-5}$ & 0.196 & 0.0117 & 0.0337 & 0.335 & 0.643 \\ 
 N300C & 0.00886 & $9.17 \times 10^{-6}$ & 0.121 & 0.0697 & 0.0200 & 0.323 & 0.720 \\
 N1600 & 0.0106 & $1.04 \times 10^{-5}$ & 0.191 & 0.00963 & 0.0312 & 0.368 & 0.608 \\ 
 N1600C & 0.0168 & $1.88 \times 10^{-5}$ & 0.272 & 0.0176 & 0.0472 & 0.353 & 0.536 \\
 N100-0.5 & 0.00310 & $1.80 \times 10^{-6}$ & 0.0987 & 0.00362 & 0.0203 & 0.292 & 0.809 \\
 N100-Z0.1 & 0.00315 & $4.74 \times 10^{-7}$ & 0.0964 & 0.00369 & 0.0269 & 0.295 & 0.806 \\
 N100-Z0.01 & 0.00316 & $1.60 \times 10^{-7}$ & 0.0947 & 0.00374 & 0.0290 & 0.290 & 0.807 \\ 
 
\hline

CCSN SNe  & & & & & & & \\
M15bE2.63 & 0.153 & 0.060 & 0.727 & 0.139 & 0.059 & 0.098 & 0.229 \\
M15bE0.92 & 0.154 & 0.060 & 0.739 & 0.142 & 0.060 & 0.091 & 0.090 \\
M15bE0.74 & 0.154 & 0.060 & 0.739 & 0.142 & 0.060 & 0.091 & 0.090 \\
M15aE0.82 & 0.154 & 0.060 & 0.741 & 0.151 & 0.062 & 0.055 & 0.021 \\ 
M15aE0.54 & 0.154 & 0.060 & 0.732 & 0.152 & 0.062 & 0.040 & 0.020 \\
M15aE2.47 & 0.154 & 0.060 & 0.727 & 0.142 & 0.059 & 0.095 & 0.197 \\ 

M20cE2.85 & 0.332 & 0.083 & 1.899 & 0.538 & 0.103 & 0.162 & 0.109 \\ 
M20cE1.65 & 0.333 & 0.083 & 1.924 & 0.540 & 0.103 & 0.161 & 0.039 \\ 
M20cE1.00 & 0.326 & 0.083 & 1.681 & 0.488 & 0.094 & 0.018 & 0.027 \\ 
M20cE2.76 & 0.332 & 0.083 & 1.898 & 0.537 & 0.103 & 0.164 & 0.089 \\ 
M20aE2.43 & 0.332 & 0.083 & 1.897 & 0.538 & 0.103 & 0.131 & 0.039 \\ 
M20bE1.04 & 0.321 & 0.083 & 1.604 & 0.457 & 0.088 & 0.018 & 0.027 \\ 
M20aE2.50 & 0.332 & 0.083 & 1.893 & 0.539 & 0.104 & 0.096 & 0.028 \\ 
M20aE0.53 & 0.283 & 0.083 & 1.006 & 0.220 & 0.048 & 0.015 & 0.026 \\

M25aE1.57 & 0.423 & 0.146 & 2.340 & 0.605 & 0.125 & 0.023 & 0.033 \\
M25d2E2.53 & 0.422 & 0.146 & 3.203 & 0.615 & 0.171 & 0.316 & 0.053 \\ 
M25d2E2.64 & 0.422 & 0.146 & 3.203 & 0.615 & 0.171 & 0.316 & 0.053 \\ 
M25d2E2.52 & 0.422 & 0.146 & 3.203 & 0.615 & 0.171 & 0.322 & 0.056 \\ 
M25d2E2.78 & 0.422 & 0.146 & 3.203 & 0.615 & 0.171 & 0.316 & 0.053 \\ 
M25d2E0.75 & 0.356 & 0.146 & 1.075 & 0.178 & 0.051 & 0.017 & 0.030 \\ 
M25d3E1.04 & 0.433 & 0.146 & 3.220 & 0.693 & 0.170 & 0.341 & 0.366 \\

\hline\hline

\end{tabular}
\caption{AGN star and field star ejecta in solar masses. The AGN yields here have contributions from stellar wind, an adiabatically expanding rotation disk wind, and ejecta composed of stellar material due to disruption of the disk wind. The C, N, O yield here do not include their modest changes
due to the CNO burning process on the main sequence track.  
The core-collapse supernova~\citep{2018ApJ...856...63F, Andrews_2020} and thermonuclear supernova \citep{2013MNRAS.429.1156S} models are used for 
comparison of final ejecta yields.}
\label{tab:yields}
\end{center}
\end{table*}

\subsection{Yields from Field Stars:  Core-collapse and Thermonuclear Supernova Yields}
\label{sec:ccyield}

To compare to our AGN yields, we must first compile the yields from field stars, both core-collapse (CCSN) and thermonuclear (TNSN) supernovae.  For core-collapse supernovae, we use the yields from 1-dimensional core-collapse calculations using energy-deposition to vary the supernova energy~\citep{2018ApJ...856...63F,Andrews_2020}.  We focus on the yields from 15, 20, and 25\,M$_\odot$ zero-age main-sequence mass stars with "normal" core-collapse supernova energies ($0.5-3 \times 10^{51} \, {\rm erg}$).  At collapse, these masses are roughly 10.7, 13.0, and 12.3\,M$_\odot$.  Table~\ref{tab:yields} shows a subset of the yields from these models showing the range of yields and, more importantly, yield ratios from these supernovae.   Our field star models eject comparable abundances of iron and silicon (Fe/Si ranges from 0.1-3 based on progenitor and explosion energy).  In our core-collapse supernovae, O dominates the yield, often 2 orders of magnitude above the yields of Si, Fe, or C.  Mg is produced at a factor of a few times lower than C.

Thermonuclear supernovae are major contributors of iron and silicon in the universe.  Because they arise from relatively low-mass progenitors, thermonuclear supernovae should be rare in AGN disks.  Stars forming in AGN disks tend to accrete mass even beyond the onset of nuclear burning in the core, driving larger masses.  For our field-star distribution, we use a broad sample of thermonuclear supernova calculations from~\cite{2013MNRAS.429.1156S}.  These results are based on 3-dimensional simulations of deflagration-initiated explosions leading to dealyed-detonation explosions.  As with most thermonuclear supernovae, they produce large amounts of $^{56}$Ni that decays to iron and this iron productions exceeds that of any other element (Fe/Si $\sim 1.5-20$).  These explosions also eject unburned carbon, oxygen and magnesium from exploding white dwarf.  But iron dominates the yields, roughly 2 orders of magnitude above the Mg and C yields, and more than an order of magnitude greater than the O yield.  Alternative explosion mechanisms (e.g. helium detonation or gravitationally-confined detonation models) produce the same trends in yield ratios~\citep{2017hsn..book.1955S}.  

\subsection{Yields from Our AGN stars}

The physics behind the formation and ejection of isotopes in our AGN stars is slightly more complex.  It includes 3 different components:  stellar winds, disk winds, and stellar disruption.  The stars produced in AGN disks reach very high masses (in excess of 400\,M$_\odot$). First generation
stars acquire pristine H and He gas (model v200lowZ) and return CO-laden wind to the disk 
gas.  Later generation star recycle disk gas which has been polluted by their predecessors
(model v200).  During their main sequence evolution,
a large fraction of the stellar mass is released with He and N abundances enhanced by the
CNO burning process \citep{2023MNRAS.526.5824A}.   
Along their post main sequence tracks, the triple-$\alpha$ reaction 
primarily produces C+O and strong winds eject H, He (typically abundances 
in field star winds), and heavier elements (with C several times larger than
O yields).  Since the wind is limited to removing the outer layers, only trace 
amounts of Ne, Mg, and Si is contained in the ejecta (Table~\ref{tab:yields}).

After the onset of collapse, non-rotating AGN stars contract into black holes and produce 
no further ejecta.  But, if these stars are rotating, a disk is formed that can undergo 
further burning and drive additional outflows.  For these models, we can calculate the 
ejecta properties from our disk models.  The initial ejecta is heated in the disk and we use the disk properties from \cite{1999ApJ...518..356P} to determine the disk temperature.  As it moves outward, it shocks the stellar material, heating it.  We assume that the wind ejecta quickly blows a jet-like funnel out through the star.  We assume the ejecta expands adiabatically, using a simple analytic disk wind model~\citep{2023ApJ...956...71K}. 
  
With these trajectories, we are able to use nuclear network codes to calculate the detailed yields. We post-process the non-interacting disk tracer particles through the NuGrid TPPNP nuclear reaction network~\citep{2015MNRAS.447.3115J,2016ApJS..225...24P}. Each trajectory is integrated in time using a variable-order Bader-Deuflhard integrator 
\citep{Bader1983-hg, Deuflhard1983-le, Timmes1999-rp}. Nuclear Statistical Equilibrium (NSE) is assumed at temperatures above 5 GK and we evolve the electron fraction via a 4th/5th order Cash-Karp type Runge–Kutta integrator (\cite{Cash1990-qc}). For more information on the NuGrid TPPNP code we refer the reader to (\cite{Jones2019-am}, \cite{Jones2019-ek}).

The conditions in the disk are typically so hot that the material is, for the most part, in nuclear statistical equilibrium.  Most of this material fuses to iron peak elements and remains into Fe as it is ejected.  The amount of iron injected into the AGN disk from these outflows depends on the fraction of the disk mass launched in the wind.  In Table~\ref{tab:yields}, we assume 20\% of the disk mass is lost through disk winds.  The disk winds tend to predominantly produce iron peak elements with only trace amounts of the lighter elements.

\begin{figure}
\centering
\includegraphics[width=0.47\textwidth]{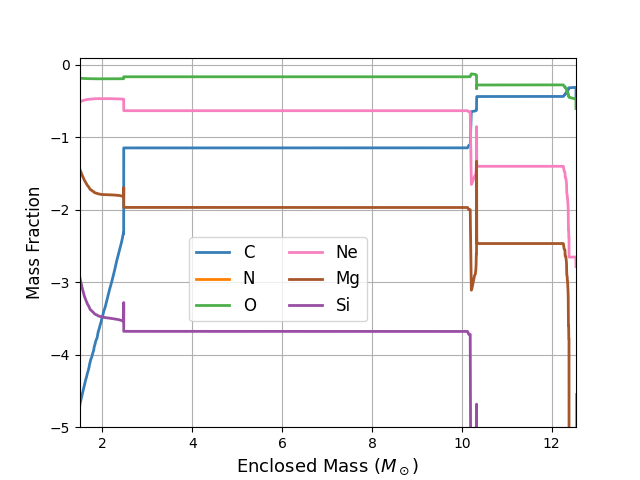}
\includegraphics[width=0.47\textwidth]{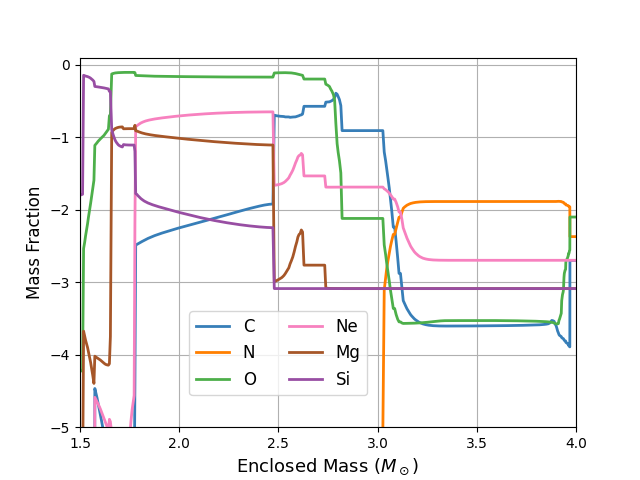}
\caption{Abundance mass fractions  of C, N, O, Ne, Mg, Si versus mass in our v200 AGN-disk star (top) and a 15\,M$_\odot$ field star (bottom). The large C/O core leads to very little variation of the abundances as a function of enclosed mass.  Most of the N is destroyed in these stars prior to collapse.  There is much more structure in our 15\,M$_\odot$ and that structure leads to very different total Mg and N abundances.}
\label{fig:abundstar}
\end{figure}

The outflow from the disk wind will also disrupt the outer envelope of the star as it clears 
out a funnel ejecting a later-arriving fraction of the stellar material.  For our stellar 
disruption yields, we calculate the stellar material exterior to the black hole when the disk 
forms and then eject a constant fraction of this exterior profile.  For Table~\ref{tab:yields}, 
we set this fraction to 20\%.  We assume that the final few days (during silicon burning) do not
significantly alter the composition of the ejecta.   This final burning phase primarily affects 
the composition of the innermost part of the star which forms the black hole.  Perhaps more 
importantly, we further assume that the shock of the disk wind propagating through the star 
does not drive further burning.  It is likely that these shocks will drive some burning that 
will slightly alter the yields from our stellar disruption.  The yields from stellar disruption 
reflect the abundances of the star at collapse, ejecting considerable C, O, and Mg.  These stars 
are very different to those of field stars.  Figure~\ref{fig:abundstar} shows the stellar 
abundances of our AGN-disk stars at collapse and those of a typical field star.  Our field 
stars produces nearly 10 times as much Mg as our AGN-disk star.  Mg is a sensitive tracer 
of the C and Ne burning shells and it is not surprising that the Mg abundance is very different 
in our peculiar AGN disk stars.  The high masses of the C/O cores in our AGN disk stars have very different structures than the 15-25\,$M_\odot$ field stars that dominate the nucleosynthetic yields in field stars.  If the C/O cores were smaller in our AGN disk stars, the yields from these stars are likely to be more similar to field stars. 

Combining all 3 components of our AGN-disk stars:  stellar winds, disk winds and stellar disruption, we note distinct differences between AGN-disk stars and field star yields.  
The low C/O ratio of our thermonuclear explosions reflects the conditions 
of the CO progenitors of these supernovae.  A similar ratio is found in the disruptive yields 
albeit most of the C+O are released before the collapse.
Our field star explosions produce Fe on par with Si.  Our AGN-disk stars produce no iron 
unless the star is rotating, but can produce considerable iron if a disk forms.  

\subsection{abundance ratios}

\begin{figure}
\centering
\includegraphics[width=0.47\textwidth]{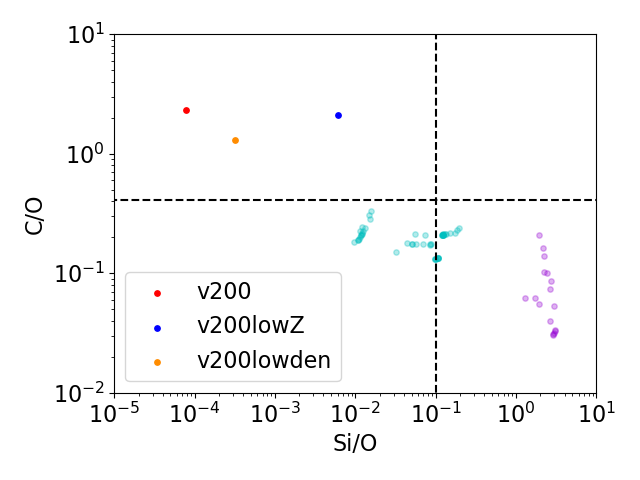}
\caption{Ejecta ratios of C to O and Si to O for our AGN models (with disk, disruption, and stellar ejecta combined), modest energy core-collapse supernova models (cyan), and thermonuclear supernova models (purple), as compared to solar ratios (black dashed line). We see here that the C/O ratio is significantly higher for our AGN models compared to field stars, while the Si/O ratio is lower than that of the field stars. We also note the Si/O ratio is in good agreement with the expectations from observations seen in \ref{fig:si_c_abund} for the low metallicity and, to a lesser extent, the low density model.}
\label{fig:oxygenratios}
\end{figure}

\begin{figure}
\centering
\includegraphics[width=0.47\textwidth]{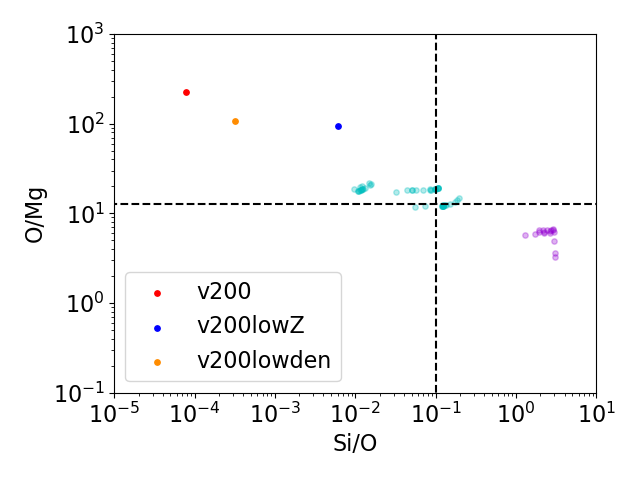}
\caption{ Ejecta ratios of O to Mg and Si to O for our AGN models, modest energy core-collapse supernova models (cyan), and thermonuclear supernova models (purple), as compared to solar ratios (black dashed line).}
\label{fig:mgratios1}
\end{figure}

\begin{figure}
\centering
\includegraphics[width=0.47\textwidth]{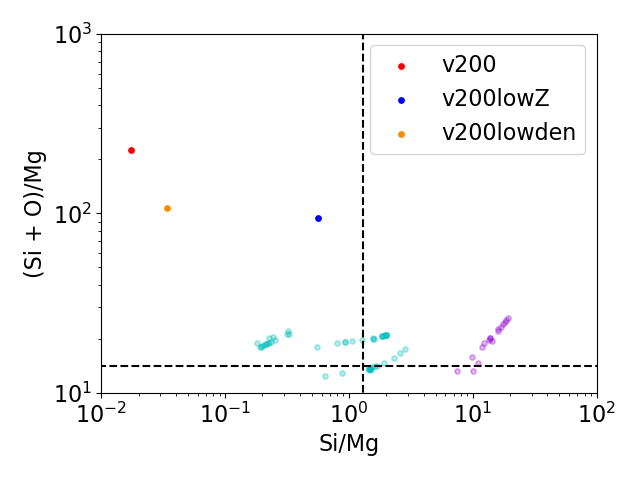}
\caption{Ejecta ratios of (Si + O) to Mg and Si to Mg for our AGN models, modest energy core-collapse supernova models (cyan), and thermonuclear supernova models (purple), as compared to solar ratios (black dashed line). Here, we do not see good agreement with Figure~\ref{fig:mg_si_abund} due to our under production of Mg.}  
\label{fig:mgratios2}
\end{figure}

\begin{figure}
\centering
\includegraphics[width=0.47\textwidth]{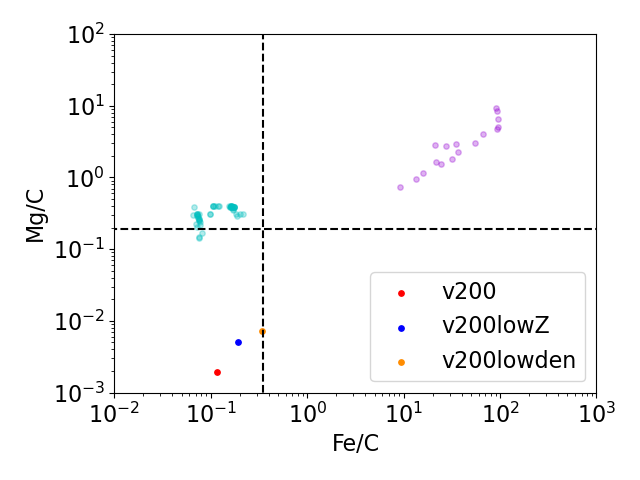}
\caption{Ejecta ratios of Mg and Fe to C for our AGN models, modest energy core-collapse supernova models (cyan), and thermonuclear supernova models (purple), as compared to solar ratios (black dashed line).}
\label{fig:Cratios2}
\end{figure}

Figures \ref{fig:oxygenratios}, \ref{fig:mgratios1}, \ref{fig:mgratios2}, and \ref{fig:Cratios2} show the abundance ratios relevant for observations presented here.   Let us compare the yields for each of these figures individually.  
In Figure~\ref{fig:oxygenratios}, we show the C/O vs. Si/O 
ratios.  The high mass in our AGN-disk stars produces a large carbon core and much of this carbon is then 
lost in the stellar winds.  The carbon yield in these stellar winds is high, leading to a C/O ratio that 
is higher than our field star yields.  
Our AGN-disk models ejected very little Si, leading to a low Si/O ratio.  However, our AGN-disk stars may 
be underestimating the size of the silicon layer (we did not model these stars to collapse).  In addition, 
if the disk wind drives a strong shock, it can also produce considerable Si.  Hence, the Si/O ratio may be 
higher for our AGN-disk models than reported here.  We shall revisit the 
Mg and Si yields with modified stellar models in follow-up investigations.

Figure~\ref{fig:mgratios1} shows the O abundance relative to Mg and Si.  The high O yield from the winds of our AGN-disk stars and relatively low Mg and Si abundances in the exploding star lead to high O/Mg ratios and low Si/O ratios with respect to supernovae.  

Figure~\ref{fig:mgratios2} shows the ratios of Si and Si+O to Mg.  The low Mg mass fractions in our AGN-disk stars relative to field stars (see Figure~\ref{fig:abundstar}) lead to very high (Si+O)/Mg ratios, but the low Si yield leads to Si/Mg ratios that are similar to core-collapse supernovae.  With the high Si yields of thermonuclear supernovae, these explosions produce the highest Si/Mg ratios.

Finally, we plot the C ratios with respect to Mg and Fe (Figure~\ref{fig:Cratios2}).  The low Mg production in our AGN-disk stars lead to low Mg/C abundances.  But, if the AGN-disk star is rotating, the high Fe production in the disk wind leads to a higher Fe/C ratio than that of normal core-collapse supernovae.  Thermonuclear supernovae which dominate Fe production produce the highest Fe/C ratios.

\section{Yields Observed in AGN Disks}
\label{sec:observ}

By comparing the calculated yields from our AGN disk stars in \S\ref{sec:simyields} to the observed abundances in AGN disks, we can distinguish between different sources for these nucleosynthetic yields.  Ultimately, these yield comparisons can determine the contribution from (and hence rate of) stars formed in AGN disks. But to do so, we must understand the composition of these disks.  More importantly, we must understand the uncertainties in these composition measurements.  

{{\subsection{Inferring the observed elemental abundances in quasar broad line region}}}
Previous work has focused on the metallicity in the quasar broad line region (BLR {, in the outskirt of the quasar accretion disc which contains the embedded stars}) relative to solar abundance. The indicator \nagao, as proposed by \cite{Nagao2006a} provides reliable estimates of BLR abundance. As the overall metallicity increases in the BLR cloud, the fraction of $\rm \cppp$ in the cloud decreases almost in inverse proportion to the abundance of heavy elements. 
This trend reflects the fact that the zone in which $\rm \cppp$ dominates gets smaller 
as the carbon abundance increases. Consequently, as metallicity increases, the intensity of the \civ\ emission line decreases while that of \siiv\ increases. Under the assumption that all the $\alpha$ elements' abundance scales with respect to the solar abundance, the inferred BLR mean metallicity of $Z\sim 3Z_{\odot}$ (e.g., \cite{hamann2002, Juarez2009}). However, to distinguish the stellar population in AGN disks, it is necessary to separately assess the abundance of individual elements.

\begin{figure}
\centering
\includegraphics[width=0.47\textwidth]{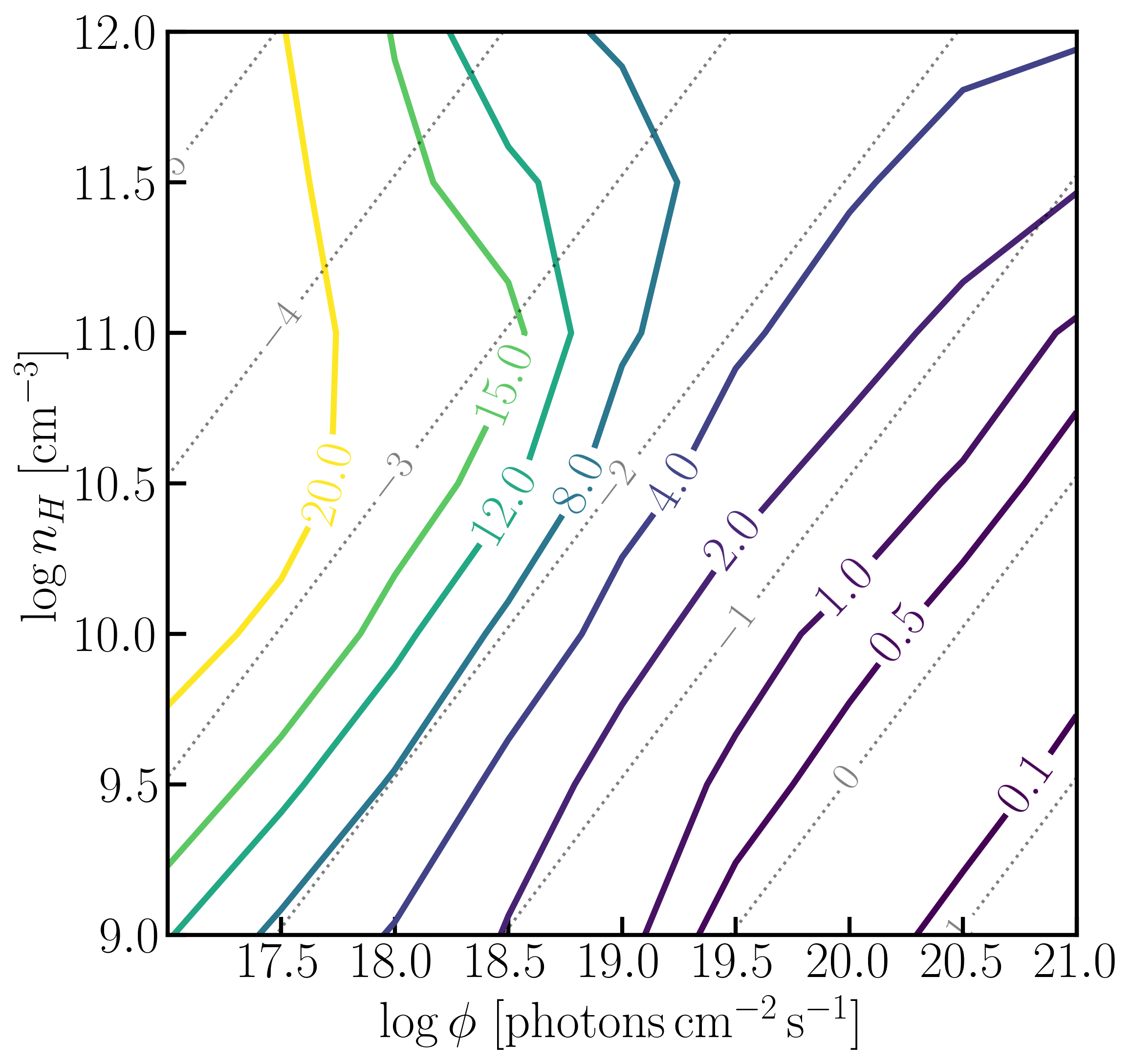}
\caption{\texttt{CLOUDY} line intensity ratio \mghb\ for $Z=3Z_{\odot}$. The colored curves correspond to different models with various densities and ionizing flux. The grey dashed contour lines show the corresponding ionization parameter, $U \propto {\rm \, ionizing \, flux}/{\rm electron \, number \, density}$. This indicates that the intensity of \mgii\ relative to \hbeta\ is a strong function $U$, where \mghb\ decreases as $U$ increase.}
\label{fig:mghb}
\end{figure}

The difficulty in using line strengths to determine the abundance fractions lies in the fact that the line strengths are determined by the excitation and ionization fractions of the elements.  We must apply models determining the line emission that depends on the conditions of the gas and the radiation impinging on this gas.  To put some constraints on the abundance of individual elements, we run a set of \texttt{CLOUDY} \citep{Chatzikos2023Cloudy} models to study the line intensity ratio for clouds with a slab geometry and the “Strong Bump” ionizing continuum employed by \cite{Nagao2006a}. We use a fixed stopping column density of $N_{\rm H} = 10^{23}$ cm$^{-2}$ \citep{Baldwin1995}, and a wide range of the ionization flux, gas density, which resembles the locally optimally emitting clouds (LOC) for BLR clouds \citep{Baldwin1995, Korista2004}. Figure \ref{fig:mghb} shows the contour of \mghb\ ratio for a fixed overall metallicity ($Z_{\alpha}=Z_{\rm Fe}=3Z_{\odot}$). For a fiducial BLR gas density of $n_{\rm H} = 10^{10} \rm cm^{-3}$ \citep{davidson1979}, the line intensity ratio decreases from \mghb\ $\sim 12$ to $\lesssim 1$ when the ionizing flux increases from $\log \phi = 18$ to $\log \phi = 20 \rm \, cm^{-2} s^{-1}$ (corresponding to an increase in the ionization parameter from $\log U = -2.5$ to  $\log U = -0.5$, where $U$ is proportional to the ionizing flux divided by the gas density). When $U$ increases, the intensity of ionizing radiation increases, causing a higher fraction of $\rm Mg^{+}$ ions to be ionized to $\rm Mg^{++}$ or higher ionization states. This decrease in the number of $\rm Mg^{+}$ ions results in a weaker \mgii\ emission line, thereby weakening its intensity relative to \hbeta. Our estimates of our abundances will depend on our choices for these properties.


One way to mitigate these uncertainties is to study line ratios.  
If the excitation and ionization states of lines from different elements are similar, the line ratios 
would be less sensitive to these conditions.  Using element ratios for similar ionization states can reduce the many orders of magnitude uncertainties due to ionizing flux to just a factor of a few.  In this section, we will discuss the uncertainties in the measurements and how they affect the abundance ratios for our comparisons shown in Table~\ref{tab:obsyields}.

{
{To summarize, our goal is to assess the chemical composition of the quasar BLR by matching the typical observed line intensity ratio to the \texttt{CLOUDY} \citep{Chatzikos2023Cloudy} model predicted line intensity ratio with model parameters suitable for the quasar BLR (see e.g. \cite{Baldwin1995, Korista2004}). To further eliminate the uncertainty induced by the variations in metallicity of individual quasars, we take composite SDSS quasars \citep{VandenBerk2001, constantin2003} as references for the observed (broad) line ratio for our analysis. We also discuss briefly the extreme values of the observed line intensity ratio among individual quasars to estimate the potential range of abundance ratio that quasar BLR can take. }

}
\begin{table}
\begin{center}
\renewcommand{\arraystretch}{1.2}
\begin{threeparttable}
\begin{tabular}{||c | c | c | c ||} 
\hline\hline              
Elements & By number & By mass & [A/B]\\
\hline
$\rm Mg/C$ & $-0.6\pm 0.1$ & $-0.3\pm 0.1$ & $0.2\pm 0.1$ \\
$\rm Si/O$ & $-2.4\pm 0.5$ & $-2.2\pm 0.5$ & $-1.2\pm 0.5$ \\
$\rm C/O$  & $-1.4\pm 0.1$ & $-1.5\pm 0.1$ & $-1.1\pm 0.1$\\
$\rm N/O$\tnote{\textasteriskcentered}  & $-0.6$ & $-0.6$ & $-0.2$\\
$\rm Fe/Mg$\tnote{\textdagger}  & $-0.9\pm 0.5$ & $-0.5\pm 0.5$ & $-0.8\pm 0.5$\\
$\rm Fe/H$  & $-3.9\pm 0.5$ & $-2.2\pm 0.5$ & $0.6\pm 0.5$\\
\hline
\end{tabular}

\begin{tablenotes}
     \item[\textasteriskcentered] The abundances are derived from a separate analysis with the N~{\sc iii]}/O~{\sc iii]} line intensity ratio in \cite{Huang2023}.
     \item[\textdagger] Derived by comparing the mean of the observed (Eddington ratio corrected, see \cite{Matsuoka2017}) \femg\ intensity ratio with the model from Figure 10 in \cite{sarkar2021}. The detailed analysis can be found in \cite{Huang2023}.
     
\end{tablenotes}
   
\caption{A summary of the observational constraints on the elemental abundance ratio in the BLR. [A/B] refers to abundance (by number) relative to the solar value. The errors are estimated only based on the observational uncertainties in the BLR emission line ratio. We emphasis that the uncertainties can only be treated as lower limits to the true error budget in the abundance ratio.}
\label{tab:obsyields}
\end{threeparttable}
\end{center}
\end{table}

\subsection{Mg to C abundance ratio}

\begin{figure}
\includegraphics[width=0.47\textwidth]{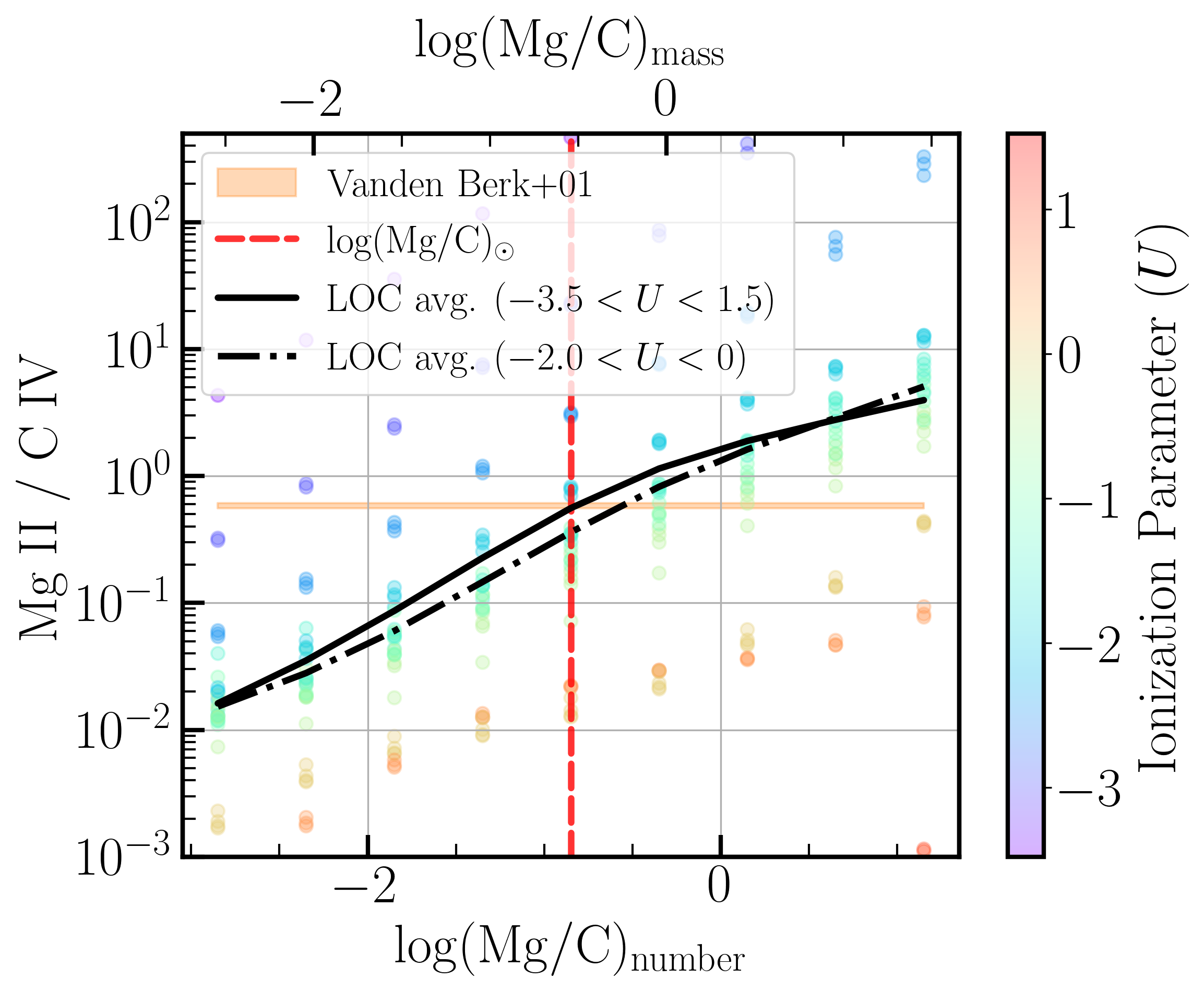} 
\includegraphics[width=0.47\textwidth]{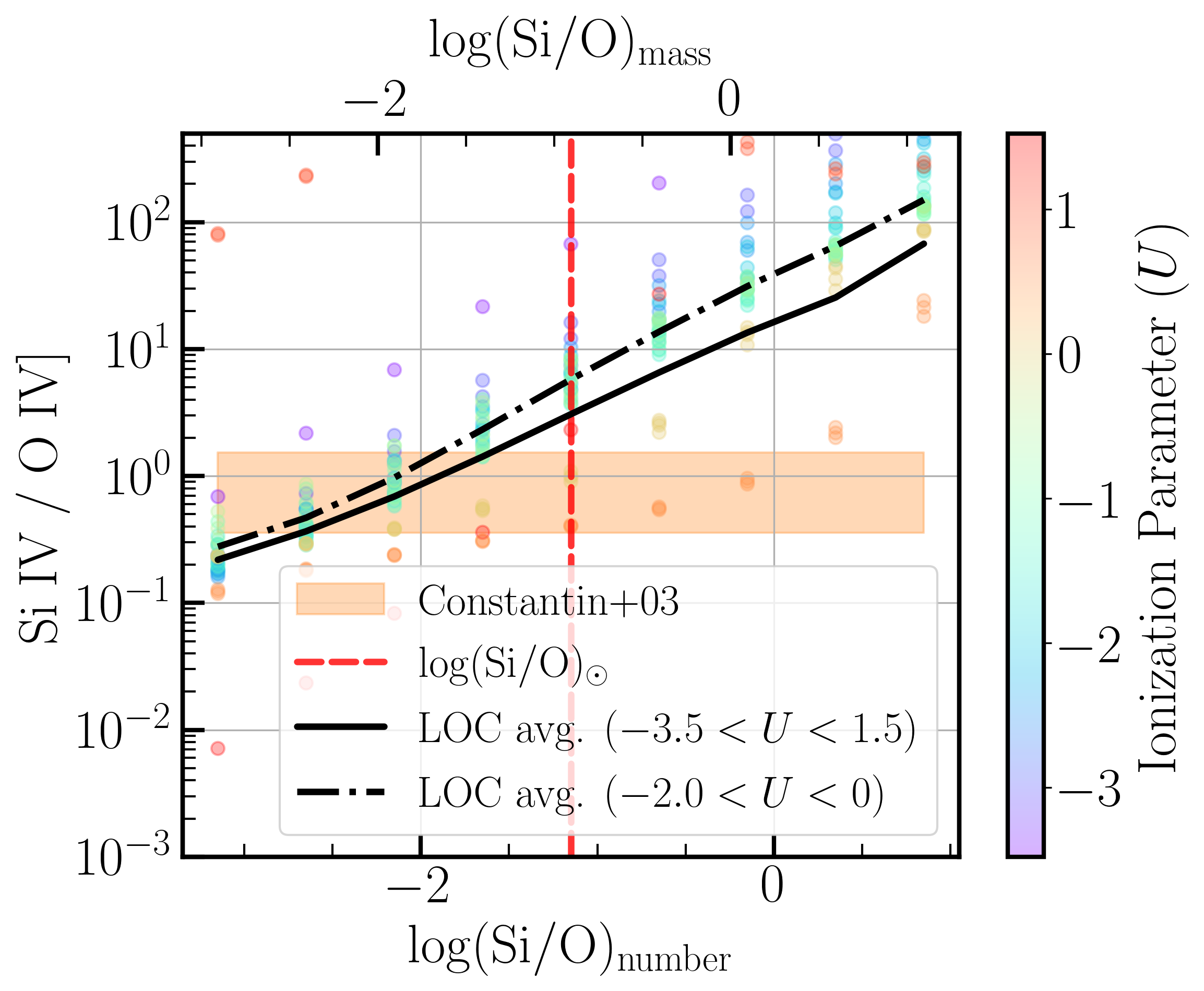}
\caption{Top Panel: \texttt{CLOUDY} line intensity ratio \mgii/\civ\ for $Z=3Z_{\odot}$. The scatter shows \mgii/\civ\ for each of the grid pairs with $18\leq \log \phi \leq 21 \, \rm cm^{-2}\,s^{-1}$ and $9\leq \log n_{\rm H} \leq 11 \, \rm cm^{-3}$. The solid black curve marks the LOC average for \mgii/\civ\ over the full range of the ionization parameter ($-3.5<U<1.5$), weighted by the \hbeta\ flux. The dotted black line shows the same line ratio, but averaged over a more realistic range of the ionization parameter for the BLR ($-2.0<U<0$). The red dashed line indicates the solar value of [Mg/C]. The observed \mgii/\civ\ ratio is $\sim 0.58$ \citep{VandenBerk2001}, as indicated by the horizontal dashed line. 
Bottom Panel: \texttt{CLOUDY} line intensity \siiv/\oiv\ ratio for $Z=3Z_{\odot}$. The orange shaded region shows the observed \siiv/\oiv\ line intensity ratio in the \citep{constantin2003} composite spectrum for the narrow-line Seyfert 1 galaxies (NLS1).}
\label{fig:mg_si_abund}
\end{figure}

The top panel of Figure~\ref{fig:mg_si_abund} compares the observed \mgii/\civ\ line ratio to the abundance ratio of Mg/C for a wide range of physical parameters ($18\leq \log \phi \leq 21 \, \rm cm^{-2}\,s^{-1}$ and $9\leq \log n \leq 11 \, \rm cm^{-3}$). The Mg/C ratio varies by as much as $\sim$ 2 dex for a fixed value of the Mg/C abundance ratio. This further confirms the strong dependence of the line intensity ratio on the ionization parameter, as seen in Figure~\ref{fig:mghb}. As a result, the difference in the \mgii/\civ\ can be attributed to the sensitivity of the low ionization line \mgii\ on the ionization parameter, $U$, as discussed in \S \ref{sec:observ}. A similar argument applies to the \civ\ intensity relative the \hbeta, where the \civ\ gets stronger relative to \hbeta\ when $U$ increases. The LOC model, on the other hand, potentially resolves the problem that both \mgii\ and \civ\ are sensitive to the ionization parameter. The LOC average, weighted by uniform \hbeta\ flux, is shown as the dash-dotted ($-2.0<U<0$) and solid ($-3.5<U<-1.5$). We focus on the narrower range of ionization parameter where $-2.0<U<0$ that brackets the physical parameter that better reproduces the bulk of the BLR spectrum. The observed line ratio, as indicated by the shaded orange region, is from the \cite{VandenBerk2001} quasar composite spectrum. The intersection between the LOC average (dash-dotted line) and the observed line ratio suggests that the Mg/C abundance ratio is around $-0.6$ by number (or $-0.3$ by mass). Given the large variation in the abundance ratio due to the variation in the ionization stage, we only attempt to estimate the lower limit for the error budget, which is given by the observed uncertainties in the emission line ratio.

\subsection{Si to O abundance ratio}
\label{Sec:Si_over_O}
\citet{Huang2023} showed that although the ratio \nagao\ can be served as a metallicity indicator for the BLR, the individual abundance ratio (e.g., Si and O) cannot be easily determined form \nagao\ ratio alone. In an attempt to disentangle the dependence on the \nagao\ ratio on the overall metallicity versus the individual abundance ratio, we run a set of \texttt{CLOUDY} models with varying the Si/O ratio and fixed overall metallicity with $Z=3Z_{\odot}$. We use the \siiv/\oiv\ line ratio as the tracer of Si/O abundance ratio. The \siiv\ is usually blended with \oiv\, which contributes extra uncertainties to the observed line ratio. Therefore, we look at Narrow-Line Seyfert 1 galaxies (NLS1), which typically exhibit narrow emission lines with FWHM(\hbeta) $\lesssim 2000 \rm km~s^{-1}$ \citep{Osterbrock1985}.
The bottom panel of Figure~\ref{fig:mg_si_abund} shows the observed \siiv/\oiv\ line intensity ratio from \cite{constantin2003} NLS1 composite spectrum. The LOC average intercepts with the observed value implies that the Si/O ratio varies from -2.9 to -1.9. A similar result is given by realistic values of the ionization parameter ($-2.0<U<0$), where Si/O ratio varies from -2.7 to -1.7.  The \oiv\ line has critical density of $n_{\rm crit} = 1.12\times 10^{11}~\rm cm^{-3}$ \citep{hamann2002}, so it is less affected by the collisional de-excitation at high density (the highest density in our \texttt{CLOUDY} model grid has $\log n_{\rm H} = 11 \, \rm cm^{-3}$). In addition, the shape of the ionizing continuum from \cite{Nagao2006a} may not accurately represent the ionizing continuum for the NLS1, due to the fact that the NLS1 galaxies typically have high Eddington ratio \citep{Boroson&Green1992}. Figure 1 in \cite{Ferland2020} shows the ionizing spectral energy distribution (SED) for AGNs with varying Eddington ratios. The peak of the ionizing continuum shifts to higher energy as we move the Eddington ratio, indicating a harder SED in the range of $\sim 30-55 \, \rm eV$, which brackets the ionizing potential of \siiv and \oiv, respectively. A direct consequence of the SED dependence on the Eddington ratio is that our LOC model average of \siiv/\oiv\ will get smaller if the bulk of the NLS1 galaxies have high Eddington ratio, therefore the inferred Si/O ratio will be higher. A final caveat for the estimation is the assumption that the underlying abundance for $\alpha$-elements is $Z=3Z_{\odot}$. This is a valid assumption based on previous studies on the BLR matellicity. However, the estimation of individual abundance ratio can be affected by the baseline abundance for $\alpha$-elements. To this end, we run the \texttt{CLOUDY} models with the same setup but with $Z=Z_{\odot}$, we find the difference between the inferred Si/O ratio to be small between $Z=Z_{\odot}$ and $Z=3Z_{\odot}$. 

\begin{figure}
\centering
\includegraphics[width=0.47\textwidth]{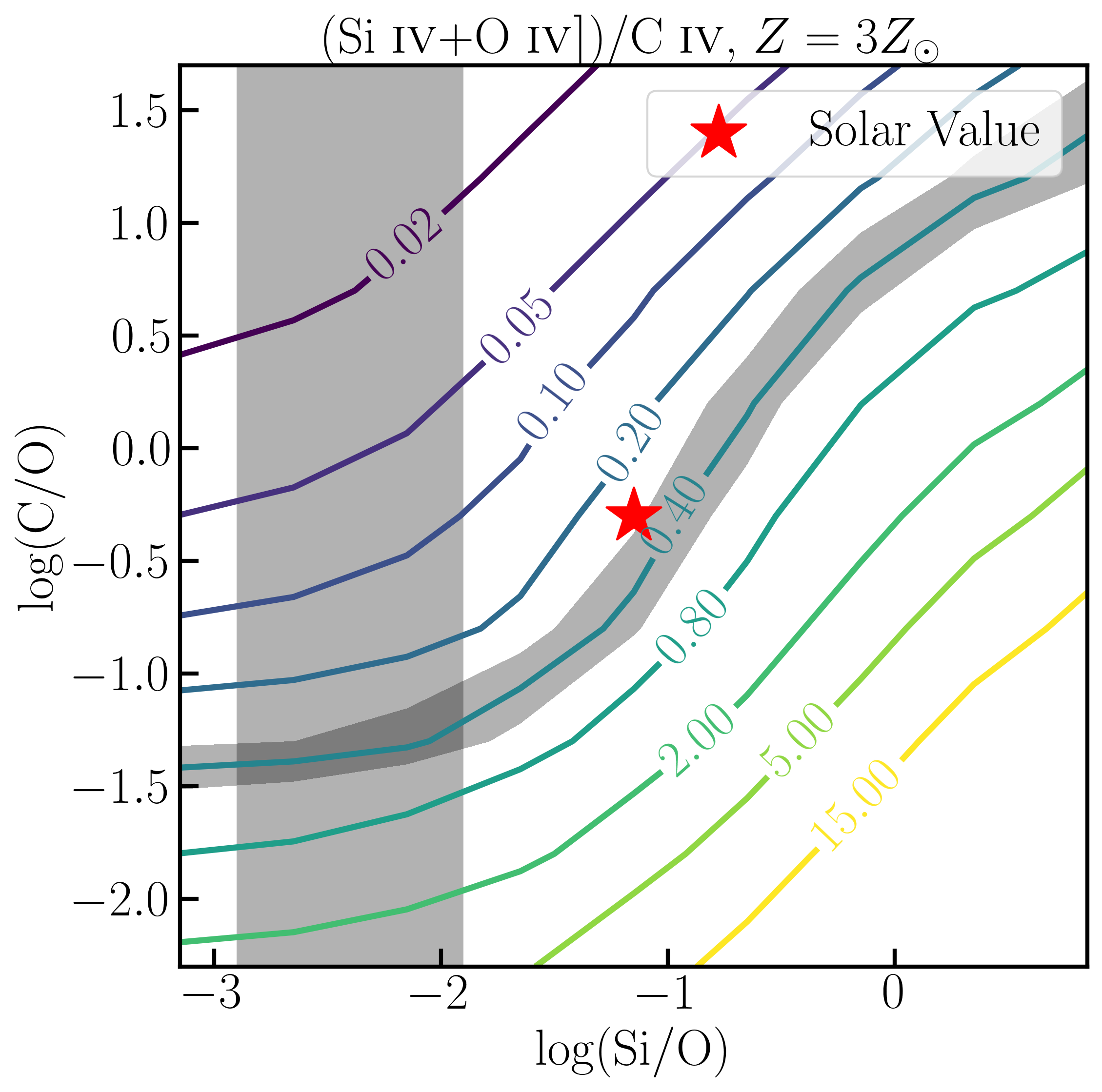}
\caption{\texttt{CLOUDY} line intensity ratio \nagao\ as a function of log(C/O) and log(Si/O) abundance ratio by number for $Z=3Z_{\odot}$.  The vertical grey shaded region marks the expected Si/O ratio (log(Si/O) = -2.9 to -1.9 by number) from comparing the observed \siiv/\oiv\ intensity ratio in \citep{constantin2003} and that of the \texttt{CLOUDY} LOC model (See Figure \ref{fig:mg_si_abund}). The grey shaded region following the contour line centered at 0.40 marks the typical observed \nagao\ ratio from SDSS quasar composite \citep{VandenBerk2001}. By construction, the intersection of the two shaded region marks the range of the inferred C/O abundance ratio, from -1.5 to -1.3. The solar abundance ratio is marked as red star. }
\label{fig:si_c_abund}
\end{figure}

As an attempt to utilize the inferred Si/O abundance to find the range of C/O abundance, we run \texttt{CLOUDY} model for the \nagao\ ratio as a function of both C/O and Si/O abundance ratio by number, where O and other $\alpha$-element abundance is kept at $\rm (O/H)=3(O/H)_{\odot}$. The result is shown in Figure \ref{fig:si_c_abund}. The vertical grey shaded region shows the value of the inferred Si/O abundance, and the shaded region following the shape of the contour at \nagao\ $\sim 0.4$ marks the typical value of the observed \nagao\ ratio.

\begin{figure}
\centering
\includegraphics[width=0.47\textwidth]{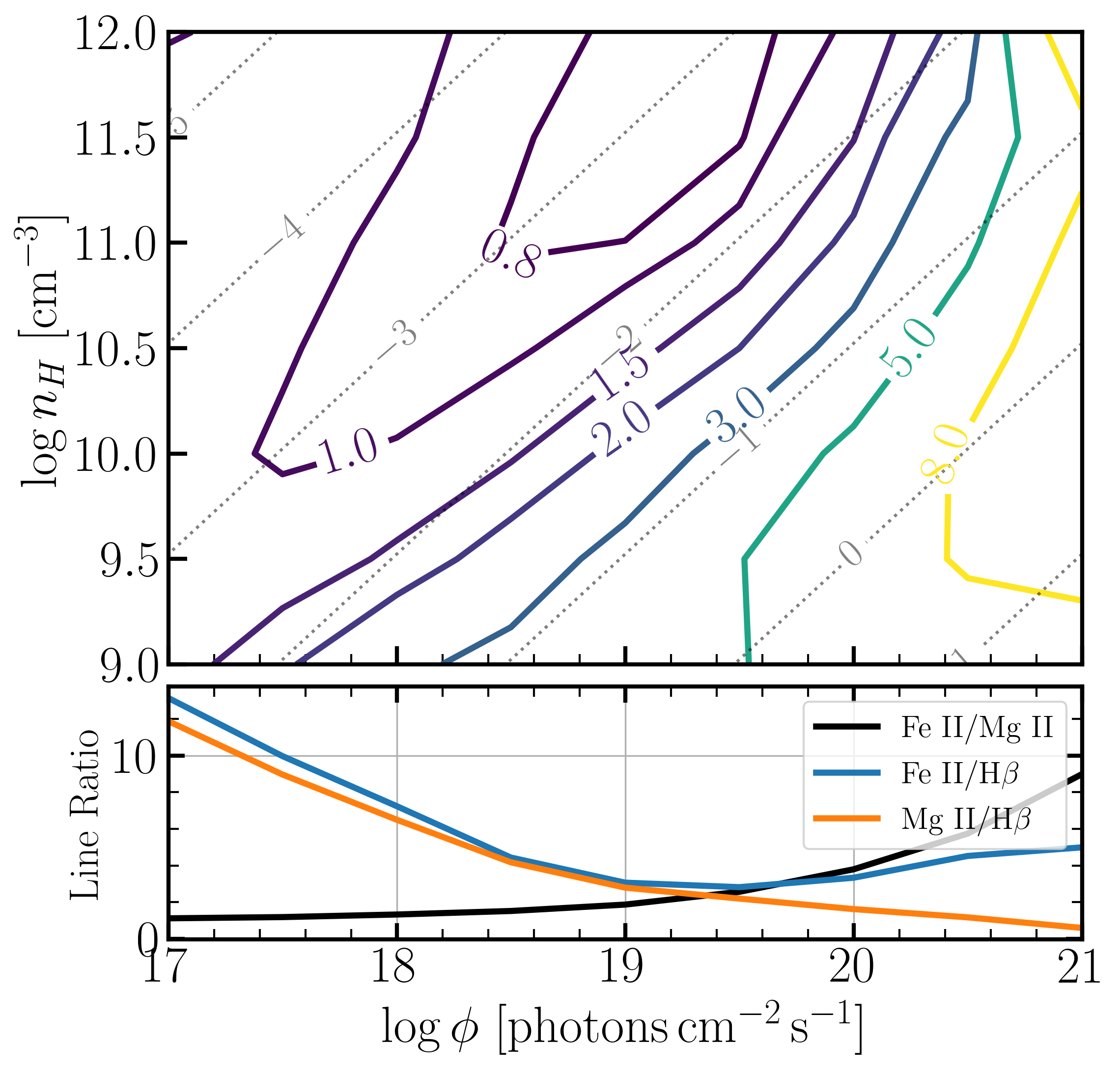}
\caption{\textbf{\textit{Top}}: Line intensity ratio \femg\ for $Z_{\alpha}=Z_{\odot}$, with a solar iron abundance. The colored curves correspond to different models with various densities and ionizing flux. The grey dashed contour lines show the corresponding ionization parameter, $U$. To first order, the \femg\ contour lines follow closely to the contour lines of the ionization parameter. This can be partially explained with the U dependency of the Mg II line equivalent width. As indicated in Figure \ref{fig:mghb}, the \mghb\ decreases as $U$ increase. \textbf{\textit{Bottom}}: line ratio \femg\ (black), \feii(UV)/\hbeta\ (blue), and  \mghb\ (orange) as a function of ionizing flux (black dashed curve). To compute the average line ratio, we assume the weight in density is uniform in $\log n_{\rm H}$, in other words, $w(n_{\rm H}) \propto n_{\rm H}^{-1}$ \citep{Baldwin1995, Korista2004}.
This demonstrates the sensitivity of the \femg\ on the ionization flux even for a fixed [Fe/Mg] abundance ratio.}
\label{fig:FeMg}
\end{figure}

\begin{figure}
\centering
\includegraphics[width=0.47\textwidth]{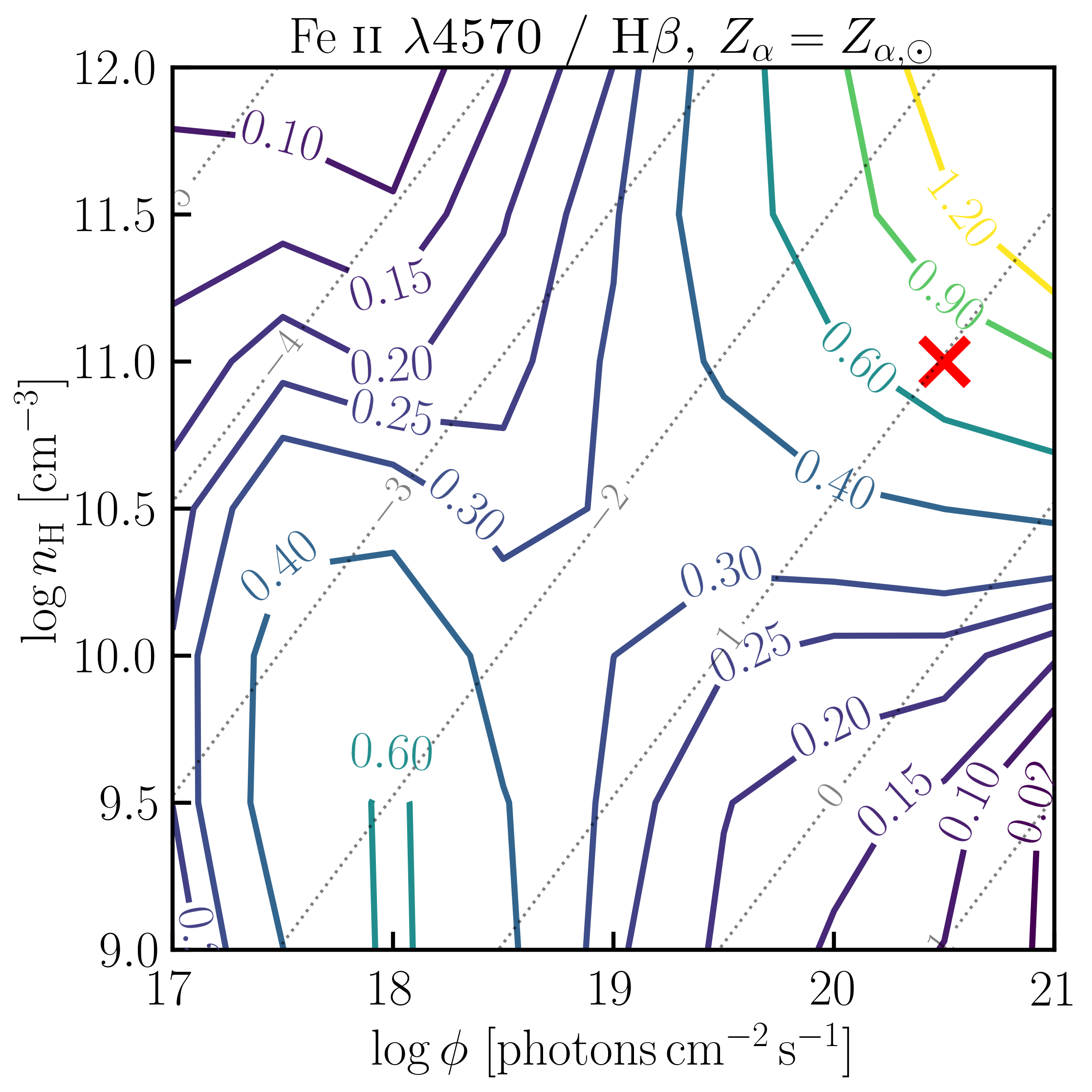}
\caption{\textbf{\textit{Top}}: Line intensity ratio \feii$\lambda 4570$ blend/\hbeta\ for $Z_{\alpha}=Z_{\odot}$ and hydrogen column density $\log N_{\rm H} = 25 \rm ~ cm^{-2}$, with super-solar iron abundance, $\rm (Fe/H) = 10(Fe/H)_{\odot}$ by number. The contour curves correspond to different models with various densities and ionizing flux. The grey dashed contour lines show the corresponding ionization parameter, $U$. The red cross marks the typical physical conditions ($\log n_{\rm H} = 11~\rm cm^{-3}$, $\log \phi = 20.5~\rm cm^{-2}\,s^{-1}$, $U=-1$) for the optical \feii-emitting region \citep{Zhang2024}. Clear enhancement in the predicted \fehb\ is seen towards high hydrogen density and large ionization flux (top-right of the contour plot), a secon.}
\label{fig:FeHb}
\end{figure}

\subsection{Constraints on the Fe/Mg and Fe/H abundance ratio}
Studies of the iron abundance in the BLR and its redshift dependence have largely focused on the ultraviolet \femg\ ratio.  
The \mgii~$\lambda 2800$ line and the UV \feii\ bands around 2400~\AA\ are close in wavelength
and involve similar stages of ionization. These studies generally find that \femg\ depends little, if at all, on redshift \citep{Maiolino2003, Verner2009, DeRosa2011, Schindler2020, Yang2021, Wang2022}.  However, caution is warranted for several reasons. For example, the \femg\ intensity ratio cannot be treated as the effect of changing iron abundance alone. As indicated on the bottom panel of Figure \ref{fig:FeMg}, with a fixed solar Fe/Mg abundance ratio, the average \femg\ intensity ratio, weighted by $n_{\rm H}^{-1}$, increases from 1.0 to 8.8 as the ionizing flux increases from $\log \phi = 17$ to $21 \, \rm cm^{-2}\,s^{-1}$. In contrast, for a single zone model ($\log n_{\rm H} = 11~\rm cm^{-3}$, $\log \phi = 20~\rm cm^{-2}\,s^{-1}$, and $v_{\rm turb}=100~\rm km~s^{-1}$), \femg\ line intensity only goes as (Fe/Mg)$^{0.2}$ \citep{sarkar2021}. Such a weak dependency of iron emission on the iron abundance implies that the uncertainty of the inferred ratio is large. As shown in Table \ref{tab:obsyields}, we assign the uncertainty in [Fe/Mg] and [Fe/H] to $\pm0.5$ only to encompass the variation in the observed line intensity ratio. If we include the uncertainty from different modeling parameters, the actual uncertainty can only be larger.

The Mg abundance in stars and stellar explosions can very dramatically with the stellar structure and explosion energies~\citep{2010NewAR..54...32M,2018ApJ...856...63F}.  Until we have run a suite of stellar models and explosive yields to determine the full range of results from our models, the Fe/Mg inferred from \femg\ may be biased by our particular small set of simulations. A potential solution is to inspect the Fe/H abundance directly from the optical \feii\ emission and its intensity relative to the \hbeta. It is worth to emphasize that many caveats arises from such an attempt: 

1) It has been well established that there exist empirical correlations between the optical \feii\ emission and fundamental AGN properties such as the SMBH mass, the Eddington ratio, the radio loudness or the X-ray properties. For example, the equivalent width ratio between optical \feii\ and \hbeta\ 
[e.g., $R_{\rm FeII} = $ EW(\feii)/EW(\hbeta) where EW(\feii) is the equivalent width of the optical blend \feii $\lambda4434-4684$] 
correlates with the width of the line FWHM (\hbeta) as well as 
the equivalent width of \oiii.  This correlation is known as the eigenvector 1 
(EV1) \citep{Boroson&Green1992, Sulentic2000}. Similar to \femg\ discussed above, the large variation in \fehb\ across objects cannot be explained by the difference in the Fe/H abundance alone. Moreover, among some changing look AGN candidates, $R_{\rm FeII}$ changes as the underlying continuum changes \citep{Gaskell2022}. \cite{Petrushevska2023} presented an changing look NLS1 where $R_{\rm FeII}$ decreases by a factor of two in $\sim 2000$ days of monitoring. A comprehensive study by \cite{Panda2022} demonstrates that the variation in the ionizing SED, together with the location where \feii\ emission is produced, can affect the intensity of the iron emission.

2) Several results from reverberation mapping have shown that the optical \feii\ emission originates in a larger radius than the typical radius of the BLR. For example, \cite{Barth2013} find that the centroid of the cross-correlation function between \feii\ and $V$-band lightcurve is $\sim 1.5-2$ times larger than that of the \hbeta\ line. This is further confirmed by the comprehensive analysis of the \feii\ emitting region in \cite{Gaskell2022}. This raises the question whether we can use a one-zone model (that is, a fixed ionizing flux and gas density) to explain the observed \fehb\ ratio.

3) As a consequence, a larger radius where optical \feii\ is emitted indicates that \feii\ emisson is be more affected by the condensation into grains. As discussed in \cite{Shields2010}, the depletions of Fe can be as large as two orders of magnitude; therefore, the abundance of Fe of the gas phase Fe abundance that we observe may not reflect the true Fe/H of the BLR gas. Our estimates can only be treated as lower limits of the true Fe abundance in the BLR.

To mitigate the above uncertainties, we propose to infer a range of possible Fe/H abundance that can help us to constrain the stellar evolution yield without trying to disentangle the physical properties that causes the variations between the optical \feii\ emission among objects. Figure \ref{fig:FeHb} shows the \feii$\lambda 4570$ blend/\hbeta\  line intensity ratio contour as a function of density and ionizing flux for $Z_{\alpha}=Z_{\odot}$ and hydrogen column density $\log N_{\rm H} = 25 \rm ~ cm^{-2}$. For a fixed ionizing SED and iron abundance $\rm [Fe/H]=1$, the \fehb\ ratio varies from $\sim 0.1$ to 1.2 depending on the ionization parameter. The typical physical condition of the optical \feii\ emitting region, $\log n_{\rm H} = 11~\rm cm^{-3}$, $\log \phi = 20.5~\rm cm^{-2}\,s^{-1}$, $U=-1$ \citep{Zhang2024}, is marked as red cross in Figure \ref{fig:FeHb}. This single-zone model gives a \fehb\ of 0.75. As a reference, the typical observed \fehb\ ratio is $\sim 0.5$, with a large variation depending on the quasar population, ranging from $\sim 0.1$ to $>1.5$. (e.g., \cite{Boroson&Green1992, Shen2014, Marziani2018}). For extreme iron emitting object such as I Zw 1, a NLS1 galaxy with \fehb\ \citep{Persson1988, Veron-Cetty2004}, \fehb\ $\sim 1.76$, the infered iron abundance may be extreme as well, if the strong \fehb\ could be caused by enhancement in high metallicity~\citep{Floris2024}, hence high [Fe/H]. On the other hand, recent discoveries of \textit{JWST} high redshift, low luminosity ($\log L_{\rm H\beta}\lesssim 44 ~ \rm {erg~s^{-1}}$) broad line AGN show weak optical \feii\ emission relative to \hbeta. {{(e.g.,  \cite{Trefoloni2024} and references therein)}. This can be explained by the low metallicity in the BLR or the depletion of Fe into grains \citep{Shields2010, Trefoloni2024}. Despite the extreme iron emitting objects, a super-solar iron abundance ($\rm [Fe/H]\gtrsim 0.6$) is needed to explain the typical observed \fehb ratio.

\subsection{Abundance Comparison}

Our fiducial AGN-star model produces $\sim 3 M_\odot$ of C and O as well as
$\sim 1 M_\odot$ of N and Fe yield  (Table \ref{tab:yields}).  
As we have indicated above, in lieu of the independent abundance determination 
for individual elements from AGNs' broad emission lines, we rely on the evaluation 
of their relative abundance and infer their abundance under the assumption
of O abundance being three times the solar value (\S \ref{Sec:Si_over_O}).  Table~\ref{tab:obsyields} 
summarizes the observed abundance ratios and their uncertainties from this 
study.  Table~\ref{tab:comp} compares the observed element ratios (by mass) 
to the ranges predicted in our AGN-disk stars, field CCSN and field TNSN models.  
The observed abundance ratios are fairly well fitted by the field CCSN models.  
The only discrepancy with the observed values is the C/O ratio from CCSN 
models being too high and the N/O ratio being too low.  

For the AGN stars, the observed N/O is much larger than that for the 
pristine v200lowZ model but comparable to that for the v200 model.  
This difference supports the assumption that the N yield mostly
comes from second or later-generation stars.  In comparison with
observation, our fiducial model over-predict C/O and under-predict
Mg/O by more than an order of magnitude.  If these fiducial models 
are correct, the high C/O ratio from AGN disk stars would place strong 
limits on the yield contributions from these stars.  
But it is worth noting that the Fe/O of the AGN-disk stars is larger than 
the observationally inferred value and that produced by CCSN.
Such high Fe yield is more than adequate to explain the large iron 
abundance fractions ($\rm [Fe/H]\gtrsim 0.6$) at high redshifts (prior 
to contributions from TNSN). 

\begin{table*}
\begin{threeparttable}
\begin{center}
\renewcommand{\arraystretch}{1.2}
\begin{tabular}{lccccc} 
\hline\hline              
Element Ratio & Observed & AGN-disk & CCSN & TNSN & Solar Value\\
\hline
$\rm Mg/C$ & $-0.3\pm 0.1$ & $<-2.0$ & $-1.0 - -0.4$ & $-0.1 - 1.0$ & $-0.55$ \\
$\rm Si/O$ & $-2.2\pm 0.5$ & $<-2.0$ & $-2.0 - -1.0$ & $-0.7 - -0.5$ & $-0.91$ \\
$\rm C/O$  & $-1.5\pm 0.1$ & $0.0-0.4$ & $-0.8- -0.2$ & $-1.5 - -0.7$ & $-0.43$ \\
$\rm N/O$  & $-0.6$ &$-4.0 - 2.0$ & $-2.0 - -1.2$ & $-6.0 - -5.0$ & $-0.82$ \\
$\rm Fe/Mg$ & $-0.5\pm 0.5$ & $\sim 2$ & $-1 - 0.6$ & $1.0 - 2.5$ & $0.28$ \\
$\rm Fe/O$\tnote{\textasteriskcentered} & $-0.4\pm 0.5$ & $-0.2 - 0.0$ & $-1.9 - -0.5$ & $0.3 - 1.7$ & $-0.69$ \\


\hline
\end{tabular}

\centering
\begin{tablenotes}
     \item[\textasteriskcentered] For better comparison with the modeled Fe yield, the abundance ratio (Fe/O) is estimated by scaling the observed (Fe/H) ratio (See Tab. \ref{tab:obsyields}) based on the assumption that $\rm (O/H)=3(O/H)_{\odot}$ by number.

\end{tablenotes}
\caption{Comparison of Abundance ratios by mass. {The observed abundance ratio is estimated by matching the observed line intensity ratio in the quasar BLR to the \texttt{CLOUDY} model predicted line intensity ratio (See \S\ref{sec:observ}).}}
\label{tab:comp}
\end{center}
\end{threeparttable}
\end{table*}

\section{Summary and discussions}
Massive stars form in AGN disks.  The continued accretion onto these stars after formation leads 
to a very different evolutionary pathways than normal field stars.  This different evolutionary 
history also produces stars whose structures at collapse differ greatly from field stars of 
the same collapse mass. The main objective of this paper is to determine the heavy-element 
(especially Fe) yield from these stars.  These quantities provide observable diagnostics 
for AGN disks' structure and place constraints on the population of remnant stellar-mass 
black holes contained within them.

In a previous analysis of this process \citep{2023MNRAS.526.5824A}, we have already
showed that 1) C+O are generated through triple-$\alpha$ reaction and released during 
the post-main-sequence evolution of all AGN stars, including the first generation stars, 
2) He and N are enriched through CNO burning and their yields are released during the 
main-sequence evolution of second or later-generation AGN stars, and 3) although 
modest amounts of Mg, Si, and Fe are generated though the $\alpha$-chain process 
on the post-main-sequence track, they are confined deeply in the core and very little 
fractions are released into the AGN disk.

Due to technical limitations, these MESA calculations end at the onset of Si 
burning. The residual cores are expected to further deplete the Mg, Si, and 
enhance Fe contents before they undergo gravitational collapse. With negligible 
spin angular momentum, the subsequent collapse of the cores with large 
residual mass ($\sim 12 M_\odot$ in the fiducial model) leads to the 
formation of stellar mass black holes without supernovae and significant
amount of Mg, Si, and Fe yields. However, 
accretion onto these post-main-sequence AGN stars causes them 
to rotate.  With a modest amount of angular momentum (corresponds to 
$\sim 0.1$ the core's break up speed), a significant 
mass fraction of the collapsing core forms a disk around a 
recently-formed central black hole.  In this paper, we show that these 
disks generate jet-driven explosions that produce large amounts of 
iron-peak elements and release ($\sim 1 M_\odot$) Fe yield into the AGN disk.  
Moreover the disk wind intercepts the delayed infall of the dwindling 
outer regions of the collapsing cores.  This interruption further modifies 
the abundance and ejects modest amounts of Mg and Si yields. 

With these collapse/explosion models, we are able to estimate the nucleosynthetic yields from AGN disk stars.  The yields are comprised of stellar wind ejecta prior to collapse, disk wind ejecta after the formation of a black hole and stellar disruption as the disk wind and disk jet plow through the star. 
These models show that, with expected rotation rates of these stars, disks form 
around the newly collapsed stellar-mass black holes.  High temperatures in these disks
lead to the prolific production of iron.  A primary result of this study is our 
demonstration that the winds from these disks can inject considerable amounts of iron into the AGN disk.  

The full nucleosynthetic yields from these fiducial models have very different 
signatures from those of field stars.  
Although our results generally support the proposition that the He, C, N, O, Si, 
and Fe abundance inferred from AGN's broad emission lines can be qualitatively 
attributed to the yields from embedded disk stars, they under-predict Mg/O and 
over-predict C/O by about an order of magnitude from, while the Si/O ratio remains within the range of observationally inferred values (Table \ref{tab:comp}).

However, both the theoretical results and the observational interpretations are still preliminary.  Our set of AGN-disk stellar (the fiducial) models uses a narrow prescription for mass accretion and subsequent mass loss.  But AGN stars form and evolve at distances ranging from $\sim 10^{-2}$ pc to a few pc from the SMBH, where the boundary conditions vary significantly.  {Although the disk properties in our study are appropriate for luminous AGNs, including quasars.  Indeed, the super-solar metallicity of the BLR has been inferred from AGNs with a range of ``Eddington ratios'' and it has no established dependence on the SMBHs' mass or on their Eddington ratio.  The presence of a population of young (S and disk) stars near Sgr A* (with a mass of $4 \times 10^6 M_\odot$) is very suggestive that in situ star formation commonly take place around AGNs with modest mass and accretion rates as well.}  

Nonetheless, in follow-up studies, we plan to 1) examine the model dependence on various parameters including the background density and sound speed of the AGN disk to determine the dependence on the disk properties and 2) modify the prescription for the radiation-pressure feedback on the accretion flow to study the dependence on our chosen mass loss model. These investigations are likely to produce a broader suite of stellar models. The rates of nucleosynthesis at various stages of AGN-stars' evolution are affected by their masses and therefore their mass accretion and wind-loss rates.  

{We employed a simplified model for the explosions produced from the stellar collapse model, focusing primarily on a parameterized mass loss.  Future work will study the energy ejection from jets and this wind mass loss and their role in driving explosive nucleosynthesis model to achieve more accurate yield and range of yield estimates.  This physics is currently poorly understood and future work, guiding by current simulations, will mostly focus on placing bounds on these yields.  Finally, in order to determine the implied abundance distribution over entire AGN disks, we also need to take into account the injection, accumulation, and diffusion of heavy elements at different AGN-disk radii.}

On the observational side, the abundance ratios for individual elements with respect to O (\S\ref{sec:observ}) includes all $\alpha$-elements (with assumed solar abundance distribution and three times the solar values), excluding those particular species. These determinations can be improved with self consistent \texttt{CLOUDY} models without the assumption of solar abundance distribution.   It is also useful to examine the time-dependent line-ratio variations among changing-look and rapidly-varying AGNs to separate the abundance contribution  from the emission-line physics. These steps are the natural extension of our quest to place  more-reliable quantitative constraints on the contribution from AGN-disk stars and to set limits on the formation rates of AGN stars and embedded black holes.  

\section*{Acknowledgements}
The work by CLF and AJA was supported by the US Department of Energy through the Los Alamos National Laboratory. Los Alamos National Laboratory is operated by Triad National Security, LLC, for the National Nuclear Security Administration of U.S.\ Department of Energy (Contract No.\ 89233218CNA000001). MAD was supported by Tamkeen under the NYU Abu
Dhabi Research Institute grant CASS. Parts of this project used python and the NumPY~\citep{2020Natur.585..357H} and Matplotlib~\citep{2007CSE.....9...90H} libraries.  We would like to thank Dr. Vera Delfavero for detailed comments and suggestions improving this paper.

\section*{Data Availability}
The data underlying this article will be shared on reasonable request to the corresponding author.

\bibliography{references} 
\bibliographystyle{mn2e}



\end{document}